\documentclass[11pt, twocolumn]{article}

\newcommand{\dgs}{$^\circ$ }
\newcommand{\dgN}{$^\circ$N}
\newcommand{\dgNs}{$^\circ$N }
\newcommand{\dgS}{$^\circ$S}
\newcommand{\dgSs}{$^\circ$S }
\newcommand{\dgW}{$^\circ$W}

\newcommand{\dgEs}{$^\circ$E }
\newcommand{\mones}{$^{-1}$ }

\newcommand{\mone}{$^{-1}$}
\newcommand{\mtwo}{$^{-2}$}

\usepackage{soul}
\usepackage{authblk}
\usepackage{natbib}
\usepackage{times}
\usepackage{mathptmx}
\usepackage{amsmath}
\usepackage{graphicx}
\usepackage[margin=0.9in]{geometry}

\title{Improving Energy-Based Estimates of Monsoon Location\\ in the Presence of Proximal Deserts\footnote{This document is a preprint of an article submitted to the AMS Journal of Climate {{http://journals.ametsoc.org/doi/abs/10.1175/JCLI-D-15-0747.1}}}}
\author[1]{Ravi Shekhar\thanks{ravi.shekhar@yale.edu}}
\author[1]{William R. Boos\thanks{william.boos@yale.edu}}
\affil[1]{Department of Geology and Geophysics, Yale University}

\date{2016 March 01}

\begin{document}
    \maketitle
    \begin{abstract}
        Two theoretical frameworks have been widely used to understand the response of monsoons to local and remote forcings:  the vertically integrated atmospheric energy budget and convective quasi-equilibrium (CQE).  Existing forms of these frameworks neglect some of the complexities of monsoons, such as the shallow meridional circulations that advect dry air from adjacent deserts into the middle and lower troposphere of monsoon regions.  Here the fidelity of energy budget and CQE theories for monsoon location is assessed in a three-dimensional beta-plane model with boundary conditions representative of an off-equatorial continent with a tropical grassland and an adjacent subtropical desert. Energy budget theories show mixed success for various SST and land surface albedo forcings, with the ITCZ being collocated with the energy flux equator but a non-monotonic relationship existing between ITCZ latitude and cross-equatorial energy transport.  Accounting for the off-equatorial position of the unperturbed energy flux equator is shown to be important when a linearization of meridional energy transports is used to quantitatively diagnose ITCZ location.  CQE theories that diagnose ITCZ location based on the subcloud moist static energy maximum are shown to have large biases; accounting for convective entrainment of  dry air by using a lower-tropospheric mean moist static energy provides a more correct diagnosis of ITCZ location.  Finally, it is shown that although ITCZ shifts can be diagnosed by modified CQE and energy budget frameworks, neither can be used in a quantitatively prognostic capacity due to unpredictable feedbacks that are often larger than the imposed forcing.
    \end{abstract}

\section {Introduction}

The Hadley circulation is one of the largest and most defining features of Earth's atmosphere, with the boundary between the two Hadley cells, the Intertropical Convergence Zone (ITCZ), containing abundant precipitating convection. The ITCZ exhibits a pronounced seasonal cycle in latitude; when it passes over land in local summer, it is associated with changes in winds and rain typically identified as monsoons. For regions on the fringes of the ITCZ's seasonal range, such as Africa's Sahel, small shifts in the ITCZ's summer latitude can bring large changes in monsoon precipitation.

The ITCZ was historically thought to be locally controlled by tropical SST and land surface properties \citep[e.g.][]{Xie2004}.  Convective quasi-equilibrium (CQE) theories for monsoon location \citep[e.g.][]{Emanuel1995, Prive2007} typically operate in this paradigm and predict that the time-mean ITCZ is constrained to lie just equatorward of the maximum subcloud layer moist static energy.  However, the ITCZ and monsoon precipitation have been shown to respond strongly to remote, high latitude thermal forcings \citep[e.g.][]{Chiang2005}, and the vertically integrated atmospheric energy budget has been used to show how ITCZ shifts are part of the anomalous meridional energy fluxes needed to balance such remote forcings \citep{Kang2008}.  This has led to two distinct yet not incompatible theories for ITCZ location -- CQE and the vertically integrated energy budget -- becoming prominent in recent years.  Yet existing forms of CQE and energy budget theories for monsoons, discussed in detail below, assume that the large-scale circulation is horizontally convergent in the lower troposphere and divergent in the upper troposphere, as in a first-baroclinic mode.  This assumption does not hold in most monsoon regions because dry, shallow circulations originate over adjacent deserts and penetrate into the precipitating monsoon domain \citep[e.g.][]{Zhang2008, Nie2010}.

This study examines how well CQE and energy budget theories diagnose ITCZ location for an idealized monsoon adjacent to a desert.  We begin, in the remainder of this introduction, by discussing details of CQE and energy budget theories for monsoon location, then review their application to the example of the West African monsoon.

\subsection{Convective quasi-equilibrium theories for monsoons}

Moist convection consumes convective available potential energy (CAPE) and restores the temperature  of the convecting layer to a vertical profile near that of a moist adiabat.  CQE theories for convectively coupled large-scale circulations \citep{Arakawa1974, Emanuel1994, Emanuel2007} assume that this process is fast compared to the generation of CAPE by buoyancy forcings, so that on longer time scales the temperature of the convecting layer covaries with the moist static energy $h$ of  air below the base of cumulus clouds.  More precisely, changes in the saturation moist static energy of the free troposphere, $h^*$, are equal to changes in the subcloud-layer moist static energy, $h_b$:  $\delta h^* \simeq \delta h_b$.  If moist convection is assumed to occupy the full depth of the troposphere, which does not seem unreasonable in the tropics, geopotential heights and thus winds are constrained through hydrostatic balance to have a first-baroclinic mode structure.

When combined with constraints on the Hadley circulation, these ideas impose a vertical structure on monsoon circulations, yielding a two-dimensional dynamical system tightly coupled to $h_b$.  Dry, nonlinear dynamical theory \citep[e.g.][]{Lindzen1988} has shown that the ascending branch of the Hadley circulation should lie just equatorward of the upper tropospheric temperature maximum; when CQE theory is used to extend this argument to a moist atmosphere, the ITCZ will then lie just equatorward of the $h_b$ maximum \citep{Emanuel1995, Prive2007}.  In the subsiding branch of the Hadley circulation, which occupies the majority of the domain, subsidence warms the free troposphere so that $h^* > h_b$ and the atmosphere is stable to moist convection.  Thus, the ITCZ should lie just equatorward of the collocated maxima in $h^*$ and $h_b$.  The circulation itself alters $h_b$ and free-tropospheric temperature through advection, so this is only a diagnostic constraint on the latitude of the ITCZ.  However, it provides mechanistic understanding, allowing one to envision how surface fluxes, radiative cooling, deep convection, and large-scale advection all interact to set $h_b$ and $h^*$.

\subsection{Energy budget theories for monsoons}

Clear demonstration that ITCZ location could be strongly affected by an imposed high latitude thermal forcing \citep{Chiang2005} opened a new line of inquiry into the cause of ITCZ shifts \citep[reviewed by][]{Chiang2012}.  Even if the location of the ITCZ is consistent with the location of the maximum in SST or $h_b$, a more fundamental mechanism must be responsible for an ITCZ shift caused by imposition of an anomalous high-latitude heat source (one could argue that midlatitude eddies transport energy to modify the tropical $h_b$ distribution, but this requires augmenting CQE frameworks with a theory for midlatitude eddies).  The vertically integrated atmospheric energy budget has become widely used in analyses of the ITCZ response to such remote forcings \citep[e.g.][]{Broccoli2006, Yoshimori2008, Kang2008}.

In the tropical atmosphere, the total energy content, which is well-approximated by $h$, tends to be slightly higher in the upper troposphere than in the lower troposphere so that the vertically integrated energy flux has the same direction as upper tropospheric winds \citep[e.g.][]{Sobel2007}.  The zonal mean, vertically integrated energy flux is thus expected to be directed away from the ITCZ, which has been called the energy flux equator (EFE): the latitude at which this energy flux is zero.  An anomalous energy sink in the high latitudes of one hemisphere would require anomalous convergence of the vertically integrated energy flux and, if there is no anomalous flux convergence in the tropics or subtropics, a meridional shift in the EFE and thus in the ITCZ toward the opposite hemisphere.  
\citet[][hereafter K08 and K09]{Kang2008, Kang2009} developed this energy budget theory for ITCZ location and showed that it quantitatively described ITCZ shifts in an idealized global model, although cloud feedbacks in their model complicated what might otherwise be a prognostic theory for ITCZ location given an imposed energy source. 

Although this energy budget theory does not impose a first-baroclinic structure on winds, as is done in strict versions of CQE via the assumption of a moist adiabat, it does assume that anomalous forcings do not change the relationship between the directions of the energy and mass fluxes.  More precisely, energy budget theories for ITCZ location assume that there are only small changes in the gross moist stability \citep[GMS;][]{Neelin1987}, which is a ratio that relates a circulation's energy transport to its mass transport.  Changes in the vertical structure of winds, temperature, or moisture that alter the efficiency of the circulation at transporting energy can produce a change in energy transports without a change in the circulation, or vice versa, which is equivalent to a change in the GMS \citep[e.g.][]{Merlis2013b}.  Although different definitions of the GMS have been proposed \citep[e.g.][]{Sobel2007, Raymond2009}, the GMS is generally positive when the time-mean circulation diverges energy away from heavily precipitating regions.  The East Pacific ITCZ is a notable example because ascent there is typically shallow, in contrast with the top-heavy ascent found in a first-baroclinic mode, and is associated with a negative GMS and a time-mean circulation that converges energy into the ITCZ \citep{Back2006, Peters2008}.

\subsection{Example of the West African monsoon}

The African Sahel lies on the poleward edge of the boreal summer ITCZ and receives most of its annual precipitation during the monsoonal ITCZ migration.  Variability in Sahel rainfall has been large on interannual and decadal time scales, with dry years associated with equatorward shifts of the ITCZ.  Early work found that Sahel rainfall variability was strongly correlated with an interhemispheric SST difference, with drought associated with colder northern and warmer southern hemisphere SSTs \citep{Folland1986, Palmer1986,Janicot1996,Giannini2003,Lu2009}.  This is qualitatively consistent with energy budget theories for monsoon location \citep{Chiang2012}, assuming that warmer SSTs are associated with  larger surface energy fluxes into the atmosphere. Sahel precipitation variability is also consistent with CQE theories, which show that the $h_b$ maximum increases in amplitude and shifts poleward during anomalously rainy seasons \citep{Eltahir1996, Hurley2013}.  Although causation is unclear, which is a drawback to the CQE framework, it is possible that SST anomalies alter the continental $h_b$ maximum through horizontal advection.

Yet there are dynamics in West Africa that are not captured by either the CQE or energy budget theories discussed above.  Sahel precipitation is correlated with surface pressure, surface air temperature, and low-level geopotential height over the Sahara in observations and in a suite of GCMs \citep{Haarsma2005, Biasutti2009}.  The circulation over the Sahara is dominated by a dry, shallow heat low circulation \citep[e.g.][]{Racz1999}, with near-surface horizontal winds converging about 1000 km north of the peak monsoon precipitation, ascent reaching to at least 3-4 km altitude, and divergence in the lower mid-troposphere \citep{Zhang2008}.  The diverging air is very dry, and one branch of this outflow from the heat low is directed toward the precipitating ITCZ at about 700 hPa.  Existing CQE and energy budget theories for monsoons do not include the effects of a proximal desert heat low and its shallow divergent circulation.  The shallow circulation is clearly not captured by a first-baroclinic mode, and outflow from the heat low could cause deviations from the moist adiabatic structure on which CQE is based.  The outflow could also suppress deep moist convection, since convection is sensitive to lower tropospheric moisture in ways not captured by the CQE assumption that $\delta h_b \simeq \delta h^*$ \citep{Derbyshire2004}.  Outflow from the heat low could  transport low $h$ air from the Sahara into the ITCZ, altering its GMS and the vertically integrated energy transports.  Indeed, \citet{Zhang2008} categorized both the Saharan heat low circulation and the bottom-heavy ascent in the East Pacific ITCZ as shallow meridional circulations, and the latter has been shown to be associated with a negative GMS \citep{Back2006, Peters2008}.  Whether vertical homogenization of $h$ by dry convection in the heat low produces a qualitatively different GMS in shallow heat low circulations is unclear.

\subsection{Goals}

Here we examine the degree to which existing CQE and energy budget theories can describe forced variations in an idealized model of a monsoon that includes an adjacent desert.  Previous studies have used idealized models of the African monsoon to assess the response to forcings, and those by \citet{Peyrille2007} and \citet{Peyrille2007a} are particularly notable.  Although \citet{Peyrille2007} did not discuss energy budget theories for monsoon location, their results are consistent with the expectation that the ITCZ will move toward an anomalous energy source and away from an anomalous sink created by SST, albedo, or aerosol forcings.  In the same model, \citet{Peyrille2007a} found that the horizontal advection of temperature and humidity accomplished by the Sahara's shallow heat low circulation played an important role in setting the latitude of the monsoonal ITCZ.  This confirms the importance of the shallow circulation in the case of the West African monsoon, motivating examination of CQE and energy budget theories in which such shallow circulations have not been considered.

Although we designed our idealized model using the West African monsoon as a reference, we expect the results to be relevant to other monsoons that lie adjacent to deserts.  Australia, southern Africa, North America, and South Asia are all regions in which a precipitating convergence zone migrates poleward to the edge of a subtropical desert during local summer, although some of these regions have less zonal symmetry than West Africa in the monsoon-desert geometry \citep[e.g.][]{Nie2010}.  
Shallow heat low circulations have been observed in all of these regions, and \citet{Trenberth2000} found twenty percent of the annual cycle variance of the global divergent mass flux to be associated with a shallow mode.  Furthermore, meridional flow often deviates from a simple first-baroclinic mode in reanalysis estimates of the zonal mean Hadley circulation \citep{Dima2003} and in idealized models \citep{Nolan2007}.  

The next section details our model configuration and methods.  Section \ref{sec:results} then shows how traditional CQE constraints fail to diagnose the location of the ITCZ and provides one possible way to restore their validity.  That section also shows that the energy flux equator and ITCZ are collocated across a wide range of forcings, but that linearizations of energy transports across the geographic equator provide a poor estimate of ITCZ location.  We close with a discussion of caveats and implications. 

\section{Methods}
\label{sec:methods}

\subsection{Model details}

The fully nonhydrostatic Weather Research and Forecasting (WRF) model, version 3.3 \citep{Skamarock2008} is used in a three dimensional configuration with 15 km horizontal resolution and 41 vertical levels between 0 and 30 km altitude. The model was modified to be on an equatorial $\beta$-plane and to have no seasonal cycle with insolation fixed at its July 15 distribution. The diurnal cycle was retained. A sponge layer was used above 20 km to absorb upward propagating gravity waves. All integrations were spun up for three months, followed by 360-day integrations (corresponding to four three-month summer seasons). Third order Runge-Kutta with adaptive timestepping is used with an average  timestep of about 40 seconds. 

We used standard WRF parametrizations of subgrid-scale physics, including the Kain-Fritsch convective parametrization \citep{Kain1990, Kain2004} and a six class bulk microphysics scheme based on \citet{Thompson2008}.  The latitude and width of the precipitation maxima are sensitive to the choice of convection scheme, but a separate integration conducted at 4 km horizontal resolution without parameterized deep convection produced a continental precipitation maximum with a similar position and width to that obtained with the Kain-Fritsch scheme (not shown).  Radiative transfer was represented by the Goddard schemes, which divide the longwave and shortwave spectra into eight and eleven bands, respectively \citep{Chou1999, Chou2001}. Vertical diffusion of heat, moisture, and momentum were handled by the Yonsei University planetary boundary layer scheme \citep{Hong2006}. 

We use a simple thermal diffusion land surface model with five layers and prescribed soil moisture, thermal inertia, albedo, and roughness \citep{Dudhia1996}.  A deep soil temperature of 287 K was imposed. Although it is somewhat unusual to use a model with prescribed soil moisture for climate studies, \citet{Xie1999} show that interactive soil moisture introduces unforced internal variability on seasonal and longer timescales in monsoons, which is especially problematic for simulations with perpetual summer insolation.  Although the effect of soil moisture on monsoons is an interesting problem in its own right \citep[e.g.][]{Douville2001, Douville2002, Koster2004}, this is peripheral to our central question of whether CQE and energy budget theories describe the time mean monsoon state.  Since there is no obvious reason to expect soil moisture-precipitation interactions to alter the relationship between ITCZ location and the distributions of $h_b$ or vertically integrated atmospheric energy fluxes, we chose to prescribe soil moisture in our perpetual-July integrations.

\begin{figure}
\centering
\includegraphics[width=19pc]{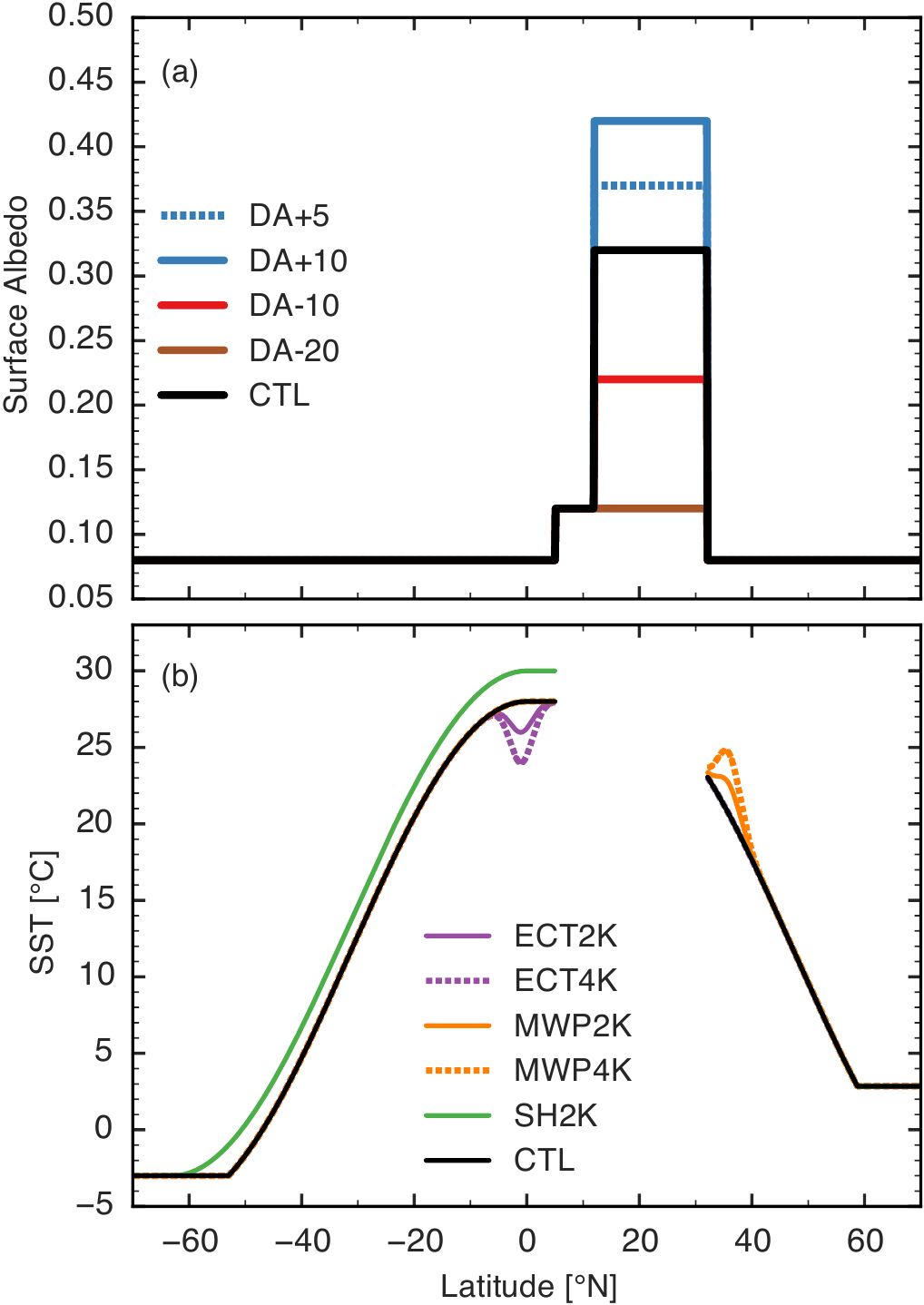}
\caption{(a) shows prescribed surface albedo and (b) shows prescribed SST. Control integration is shown in black, and applied forcings are shown in colors, according to legend. }
\label{fig:expsetup}
\end{figure}

We prescribed a quasi-sinusoidal SST (Fig.~\ref{fig:expsetup}), which was chosen to approximate average boreal summer SST. It is an adjusted version of the ``Control-5N'' profile from \citet{Neale2000}, modified to reach a maximum of 28$^{\circ}$C from 2.5\dgSs to 5\dgN. Previous studies of the relation between ITCZ location and vertically integrated atmospheric energy fluxes typically used interactive SST, most commonly with a dynamically passive slab ocean (e.g.\ K08). \citet{Kang2012a} compared the results of simulations with fixed SST and a slab ocean in detail. They illustrated how the energy budget in simulations with a slab ocean is entirely determined by top of atmosphere radiative fluxes, whereas use of fixed SST introduces surface energy fluxes that imply subsurface sources and sinks of energy. These implied sources and sinks need not balance in the global mean and can result in differing ITCZ shifts in fixed SST simulations as compared to slab ocean simulations when the same forcing is applied. However, using fixed SST allows us to crudely represent energy tendencies associated with dynamical ocean heat flux convergence, and we show this may be particularly important in near-equatorial regions in our model and in observations. Furthermore, atmospheric energy transports must still be consistent with the Hadley circulation, and there is no obvious reason to believe that the relationship between the ITCZ and the EFE would change because of the use of prescribed SST, which is consistent with the results of \citet{Kang2012a}. 

Vegetation and land surface albedo in West Africa have a large degree of zonal symmetry, and although there is some topography, the continent is relatively flat.  So we specify zonally symmetric lower boundary conditions in an equatorial $\beta$-plane spanning 70\dgS-70\dgNs and 10\dgW-10\dgEs (the domain has 1038 meridional and 149 zonal grid points at 15 km resolution). We used periodic zonal boundary conditions and closed meridional boundary conditions. Ocean exists south of 5\dgNs and north of 32\dgN. From 5\dgN-12\dgN, there is a grassland with standard United States Geological Survey (USGS) surface properties  specified in the WRF model. From 12\dgN-32\dgN, USGS surface properties for a desert were specified. We modified the desert albedo to be 0.32, which is roughly similar to the albedo of the Sahara. We also reduced desert soil moisture to the atypically low value of 1 kg m$^{-3}$; higher values produced large precipitation over the desert because of our use of specified soil moisture and perpetual summer insolation. This choice produces reasonable sensible and latent heat fluxes over the desert when compared with reanalysis data. The entire continent was set to an elevation of one meter above sea level. 

We conduct a set of integrations (Table~\ref{tab:experiments}) in which we force the model with either anomalous prescribed albedo or anomalous prescribed SST.  A subset of these forcings is illustrated in Fig.~\ref{fig:expsetup}, where the SST and albedo for the control (CTL) integration are shown in black.  Integrations designated as DA$\pm\alpha$ have the broadband shortwave \emph{desert land surface albedo} increased or decreased by $\alpha$ relative to the CTL integration, with $\alpha$ expressed as a percentage.  SST anomalies that are Gaussian in latitude with standard deviation 2\dgs and amplitude of -2 K and +2 K are applied in \emph{equatorial cold tongue} (ECT2K) and \emph{midlatitude warm pool} (MWP2K) integrations, respectively. Additional integrations were conducted with the amplitude of these anomalies set to 4 K. The SH2K integration uses SST increased by 2 K south of the continent, with most of this SST perturbation lying in the \emph{southern hemisphere}. Finally, GRASS imposes a grassland across the whole continent, from 5\dgN-32\dgN, and is equivalent to DA-20 but with increased prescribed soil moisture from 12\dgN-32\dgN.  Although some of these integrations test states are very different from the current state of West Africa, they provide a test of energy budget and CQE diagnostics of ITCZ location.  Some integrations excluded from Fig.~\ref{fig:expsetup} only apply an albedo forcing over the northern (22\dgN-32\dgN) or southern (12\dgN-22\dgN) half of the desert. Others apply more than one forcing: CEG2K applies a \emph{cross equatorial gradient} of 2 K by superimposing the ECT2K and MWP2K forcings, and AFRICA superimposes ECT2K and DA+10 forcings.  The AFRICA integration is so named because its climate most closely resembles that observed over West Africa. 

\begin{table*}
\centering
\begin{tabular}[htb]{{l|l}}
\hline
Integration & Applied Forcing from CTL\\\hline
CTL & None\\
DA+10 & Desert Albedo + 0.10\\
DA+10\_EQ & Desert Albedo + 0.10 from 12\dgNs to 22\dgNs\\
DA+5 & Desert Albedo + 0.05\\
DA-10 & Desert Albedo - 0.10\\
DA-10\_EQ & Desert Albedo - 0.10 from 12\dgNs to 22\dgNs\\
DA-10\_POL & Desert Albedo - 0.10 from 22\dgNs to 32\dgNs\\
DA-20 & Desert Albedo - 0.20\\
ECT2K (ECT4K) & -2 K (-4 K) Equatorial cold tongue\\
GRASS & Grassland across entire continent. No desert. \\
MWP2K (MWP4K) & +2 K (+4 K) Midlatitude warm pool\\
SH2K & +2 K SST Anomaly south of 5\dgNs\\
\hline
AFRICA & ECT2K and DA+10 \\
CEG2K & ECT2K and MWP2K \\
\hline
\end{tabular}
\caption{Model integrations performed and their respective forcing(s).}
\label{tab:experiments}
\end{table*}

\subsection{Convective quasi-equilibrium metrics}

To formalize CQE predictions of ITCZ location, we define a pair of estimators (borrowing terminology from statistics) of ITCZ latitude. We denote estimates of ITCZ latitude as $\hat\varphi$ and measured latitudes as $\phi$. The first estimator, $\hat\varphi_\mathrm{CQE:BL}$ is the latitude of the maximum subcloud moist static energy,
\begin{equation}
    \hat \varphi_\mathrm{CQE:BL} = \left.\phi\right|\max(h_b(\phi)) \label{equ:cqebl}\\
\end{equation}
where $h_b$ is averaged from 20 to 40 hPa above the surface. Consistent with past literature on CQE \citep[e.g.][]{Arakawa1974, Emanuel1994, Nie2010}, this layer was chosen to be below the cloud base, which in our study and in observations is typically 50~hPa above the surface \citep{Garstang1974}. We discuss the sensitivity of our averages to the thickness of this layer in the Results section.  With some foresight about our findings, we also define an estimator based on $h_\mathrm{lower}$, the moist static energy averaged over the lower troposphere, from 20 hPa above the surface to 500 hPa.  The latitude of the maximum value of $h_\mathrm{lower}$ constitutes our deep layer CQE estimate, $\hat\varphi_\mathrm{CQE:DEEP}$,
\begin{equation}
    \hat \varphi_\mathrm{CQE:DEEP} = \left.\phi\right|\max(h_\mathrm{lower}(\phi)).
    \label{equ:cqedeep}
\end{equation}

\subsection{Energy budget metrics}
\label{sec:msebudget}

Although previous studies have calculated the vertically integrated moist static energy budget in similar  contexts (e.g.\ K08, K09), we detail our approach here because of important subtleties in methodology.  We loosely follow \citet{Sobel2007} and define the mass weighted column integral of a quantity $\Lambda$ from the surface to the model top as $\left<\Lambda \right> = - \int_{p_s}^{p_t} \Lambda\;dp/g$. The time mean (denoted by an overbar) column integrated $h$ budget becomes
\begin{equation}
  \overline{\left<\partial_t h\right>} +
  \overline{\left<\mathbf{v} \cdot \nabla h \right>}+
  \overline{\left<\omega \partial_p h\right>}=
  \overline{\left<Q\right>}
  \label{equ:budget3}
\end{equation}
The first term is the atmospheric storage (or time tendency) term, the second is horizontal advection with $\nabla$ the horizontal gradient operator, and the third is vertical advection.  The fourth term is the column source of $h$, which can be decomposed as $\overline{\left< Q \right>} = \overline H + \overline E + \overline{\left< R \right>}$, with surface sensible heat flux $H$, surface latent heat flux (evaporation) $E$, and column integrated radiative heating $\left< R \right>$. By convention, energy fluxes into the column are defined as positive. Calculation of vertically integrated energy budgets can be sensitive to subtle numerical and methodological errors, such as those introduced by inadequate temporal sampling or regridding to pressure coordinates \citep{Trenberth2002}. We perform all calculations in model coordinates, transforming all equations presented below as described in the Appendix.

We begin by dropping the storage term, which in our model is indistinguishable from zero ($\approx$ 0.03 $\pm$ 0.04~Wm$^{-2}$), consistent with our use of perpetual July insolation and prescribed SST.  The two advection terms in (\ref{equ:budget3}) can be decomposed into advection by the time and zonal-mean flow and advection by eddies. Unsurprisingly, our zonally symmetric and periodic boundary conditions produce negligible stationary eddies ($\approx$ 0.5~W m$^{-2}$ in column-integrated advection), so nearly all eddy advection is due to transient eddies.  Representing deviations from the time mean by primes, (\ref{equ:budget3}) can then be rewritten
\begin{equation}
  \left<\overline{\mathbf{v}} \cdot \nabla \overline h \right>+
  \left<\overline{\omega} \partial_p\overline h \right> +
  \overline{\left<\mathbf{v} ^\prime \cdot \nabla h ^\prime \right> } +
  \overline{\left<\omega ^\prime\partial_p h^\prime \right> }=
  \overline{\left<Q\right>}
  \label{equ:budget4}
\end{equation}
Following \citet{Peters2008}, we denote the mean flow horizontal and vertical advection as $\mathrm{HADVH}=-\left<\overline{\mathbf{v}} \cdot \nabla \overline h\right>$ and $\mathrm{VADVH} = -\left<\overline{\omega} \partial_p\overline h \right>$, respectively, with their sum being the mean advection MADVH. The total eddy advection is $\mathrm{EADVH} = -\overline{\left<\mathbf{v} ^\prime \cdot \nabla h ^\prime \right>}-\overline{\left<\omega ^\prime\partial_p h^\prime \right>}$. The sum of mean advection (MADVH) and eddy advection (EADVH) is the total advection of $h$ (TADVH). 

In the time mean, neglecting storage, nonzero TADVH implies a divergent vertically integrated atmospheric energy transport (AET),
\begin{equation}
  \mathrm{AET}(\phi) = -\int_{x_{min}}^{x_{max}} \int_{\phi_{min}}^\phi \mathrm{TADVH}(x, \phi^\prime)\,d\phi^\prime dx,
  \label{equ:AET}
\end{equation}
where AET is positive if the circulation transports $h$ toward the north pole, and the latitude of the tropical zero in AET is the latitude of the energy flux equator (EFE). We express AET in petawatts and multiply it by a factor of 18 (arising from the model width of 20\dgs longitude) to ease comparison with analogous quantities observed on Earth. Using $\overline{\left< Q \right>}$ instead of TADVH in (\ref{equ:AET}) produces highly similar results (see Appendix).  This finally allows definition of the energy budget estimator for the ITCZ, which is the latitude of the most poleward zero value of AET in the tropics,
\begin{equation}
    \hat \varphi_\mathrm{EFE} = \left.\phi\right| \left[ \mathrm{AET}(\phi)=0 \right]
    \label{equ:efeestimator}
\end{equation}

\citet[][hereafter BS14]{Bischoff2014} showed that ITCZ location is related to the amount of energy transported across the equator by the Hadley circulation and to the tropical-mean $\overline{\left< Q \right>}$. Using a series expansion of AET about the equator, they showed that EFE latitude can be approximated by
\begin{equation}
  \hat \varphi_\mathrm{BS14:EQ} =  \frac{- 1}{a}\frac{
  \mathrm{AET(\phi=0)}}{\mathrm{\overline Q}_\mathrm{trop}}
  \label{equ:BS14}
\end{equation}
where $a$ is Earth's radius, AET($\phi=0$) is the cross equatorial AET, and $\mathrm{\overline Q}_\mathrm{trop}$ is the time mean, column integrated, tropical (30\dgS-30\dgN) mean $\overline{\left<Q\right>}$. Although this is slightly different from the choice by BS14 to evaluate $\overline{\left<Q\right>}$ at the equator, it prevents division by near-zero and negative values of $\overline{\left<Q\right>}$, which we show below occur in our model and in observations. 
%, typically between 20 and 30 Wm$^{-2}$.
Implicit in this scaling is the assumption that AET is a monotonic function of latitude between the equator and the ITCZ.  More generally, the series expansions can be taken about any latitude, and with foresight we define a second estimator using the same linearization about $E_0$, the latitude of the basic state EFE,
\begin{equation}
  \hat \varphi_\mathrm{BS14:E_0} =  \frac{- 1}{a}\frac{
  \mathrm{AET}(\phi=E_0)}{\mathrm{\overline Q}_\mathrm{trop}} + E_0
  \label{equ:BS14mod}
\end{equation}

\section{Results}
\label{sec:results}

\subsection{Basic state circulation}
\label{sec:ctlexp}

\begin{figure}
    \centering
    \includegraphics[width=19pc]{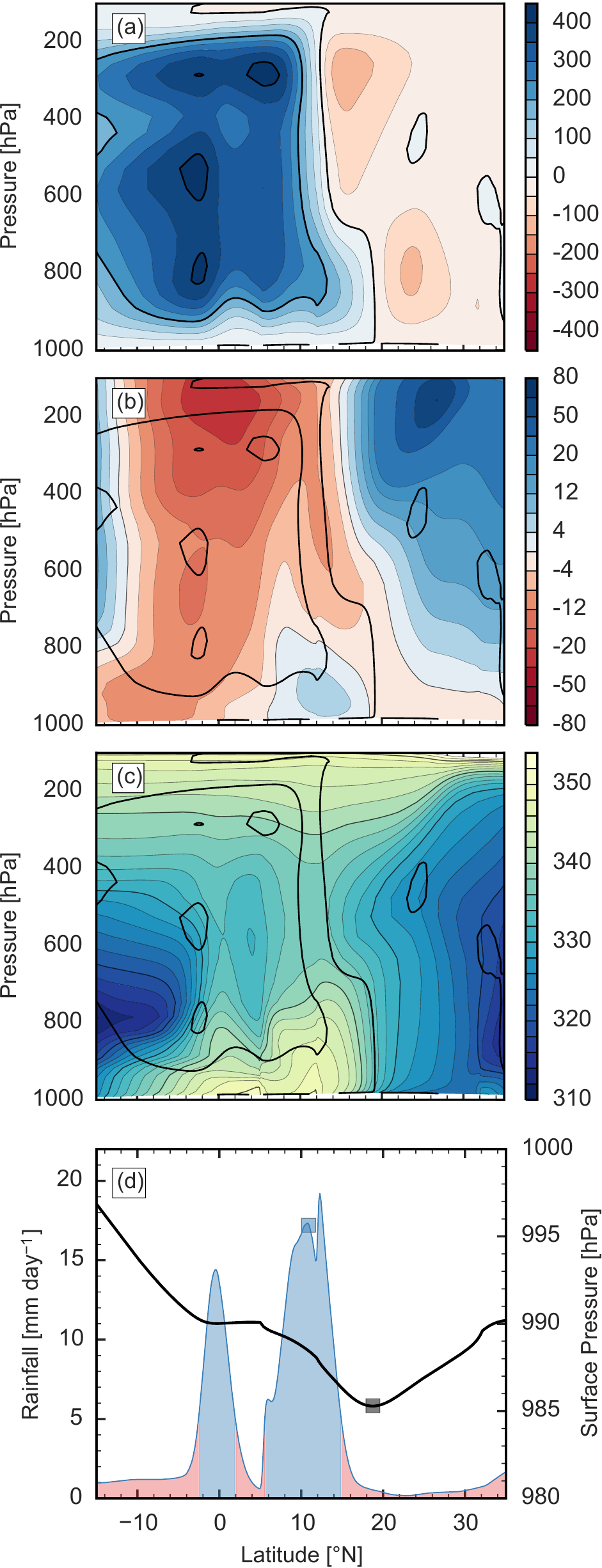}
    \caption{Zonal and time averages. (a) Mass streamfunction ($\Psi$) in units of 10$^9$ kg s$^{-1}$. Multiples of 200 have thick contours. (b) Zonal Wind (m s$^{-1}$), with $\Psi$ contours for reference. (c) Moist static energy (kJ kg$^{-1}$) also with $\Psi$ contours. (d) Precipitation (left axis; shaded) and surface pressure (right axis; black). Precipitation is shaded blue above 5 mm day\mone and red otherwise. $\phi_\mathrm{ITCZ}$ and $\phi_\mathrm{HL}$ are marked with squares. }
    \label{fig:sfnuh}
\end{figure}

The time and zonal mean mass streamfunction in the control integration consists of a deep, cross-equatorial Hadley cell with its ascent branch centered near 11\dgN, superimposed on a shallow meridional circulation with an ascent branch centered between 15-20\dgNs (Fig~\ref{fig:sfnuh}a). The streamfunction was calculated using the method of \citet{Doos2011}, which includes the effects of fluctuations in the surface pressure, a small but important contribution over the desert that prevents nonzero contours from intersecting the ground.  The winter Hadley cell in the control integration extends from 20\dgSs to 11\dgNs and is stronger than expected from reanalyses \citep[e.g.][]{Mitas2005,Merlis2013}; this is unsurprising because the off-equatorial continent occupies the entire zonal extent of the model. The summer Hadley cell is visible primarily in the upper troposphere between 10-30\dgN, with the heat low's shallow meridional circulation dominating between the surface and $\sim$650 hPa at those latitudes.  Much of the time-mean outflow from the heat low is directed toward the ITCZ, advecting hot, dry air into that precipitating region.  Other deviations from a first-baroclinic mode structure are seen, with additional inflow to and outflow from the ITCZ occurring near 350 hPa and 450 hPa, respectively, consistent with the multilevel flows found by \citet{Nolan2010} in observations and in models. 

A mid-level easterly jet is centered at $\sim$500 hPa with core velocity of 12 m s$^{-1}$ (Fig.~\ref{fig:sfnuh}b).  This is analogous to the African Easterly jet (AEJ) and, although it is positioned at a slightly higher altitude than in observations ($\sim$600 hPa), is similar in strength and location \citep{Nicholson2013}. Separate model integrations showed that the AEJ was unrealistically strong when the domain width was decreased below 10\dgs  longitude, seemingly due to insufficient momentum transports by barotropic-baroclinic eddies.  A low level off-equatorial westerly jet is centered near the ITCZ, as expected in a monsoon. 

The rainfall distribution shows two peaks (Fig.~\ref{fig:sfnuh}d), one associated with the ITCZ around 11\dgN, and a secondary peak just south of the equator associated with an equatorial ``jump'' in the zonal mean mass streamfunction. \citet{Pauluis2004} explores the equatorial jump in detail, arguing that because the Coriolis force is small near the equator, friction must be balanced by a low level cross equatorial pressure gradient. In the absence of such a gradient,  air cannot cross the equator in the boundary layer and must ascend to cross in the free troposphere. Ascent in the equatorial jump seems to be deeper in models than in observations, as evidenced by the frequency of occurence of double ITCZs \citep[e.g][]{Prive2007, Zhang2001}, and our model is no exception. We are interested primarily in the continental monsoon precipitation distribution, and define $\phi_\mathrm{ITCZ}$ as the ``centroid'' of the zonally averaged precipitation distribution over the continent from 5\dgNs to 32\dgN:  half of the spatially averaged continental precipitation occurs south of $\phi_\mathrm{ITCZ}$ and half occurs north. We use this definition instead of a simple latitude of maximum precipitation due to its superior sensitivity to subtle shifts in the precipitation distribution, as suggested in \citet[][hereafer D13]{Donohoe2013}. Consistent with observations of West Africa, the surface pressure minimum, which we define as $\phi_\mathrm{HL}$, is at the center of the heat low, about 900 km north of the precipitating ITCZ. 

\subsection{Convective quasi-equilibrium evaluation}
\label{sec:resultCQE}

\begin{figure}
    \centering
    \includegraphics[width=19pc]{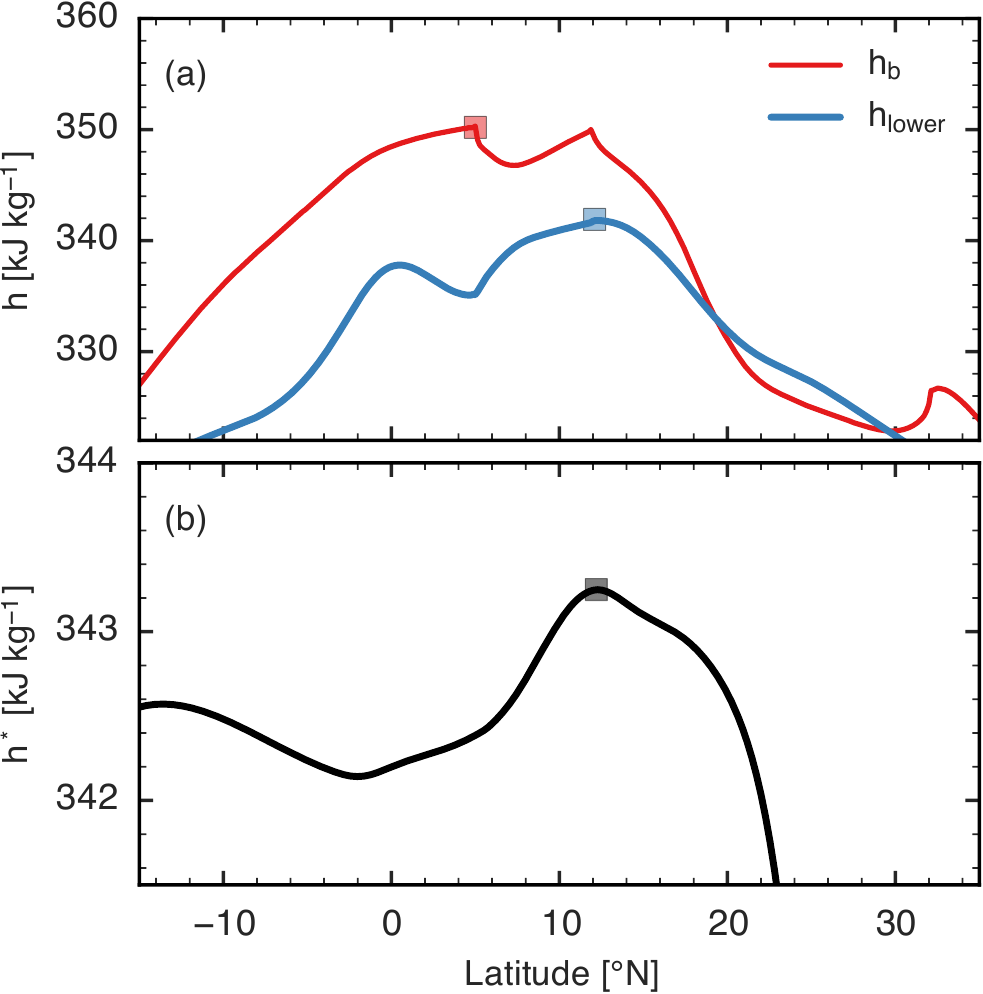}
    \caption{(a) $h_b$ (red), $h_\mathrm{lower}$ (blue) and (b) $h^*$.
    All global maxima marked.}
    \label{fig:cqeeb}
\end{figure}

In CQE theories of monsoons \citep[e.g.][]{Emanuel1995, Prive2007}, the ITCZ lies just on the equatorial side of the maximum subcloud moist static energy, $h_b$, but in our control integration there are two $h_b$ maxima of nearly equal amplitude, at 5\dgNs and 12\dgNs (Fig.~\ref{fig:cqeeb}a).  This feature persists regardless of whether the daily maximum $h_b$ \citep[e.g.][]{Nie2010} or time mean $h_b$ is used.  The $h_b$ maximum at 12\dgNs  is coincident with the maximum continental precipitation, but the maximum at 5\dgNs lies in a region of nearly zero precipitation between the continental and equatorial precipitation peaks.  The maximum at 5\dgNs is slightly stronger, leading to the estimator $\hat\varphi_\mathrm{CQE:BL}$ being located at 5\dgN, far on the equatorial side of the continental ITCZ.  Dual maxima in $h_b$ are not seen in the observed West African monsoon, at least in the comparatively coarse-resolution data provided by reanalyses, and it is possible that the bimodal $h_b$ distribution is an artifact of the fixed soil moisture or sharp transitions in surface type imposed in our model.  However, the climatological mean $h_b$ distribution over eastern Africa is in fact bimodal during July \citep{Nie2010}; this, together with theoretical interest in our model's mean state, motivates closer examination of this simulated thermodynamic distribution.

\begin{figure}
    \centering
    \includegraphics[width=19pc]{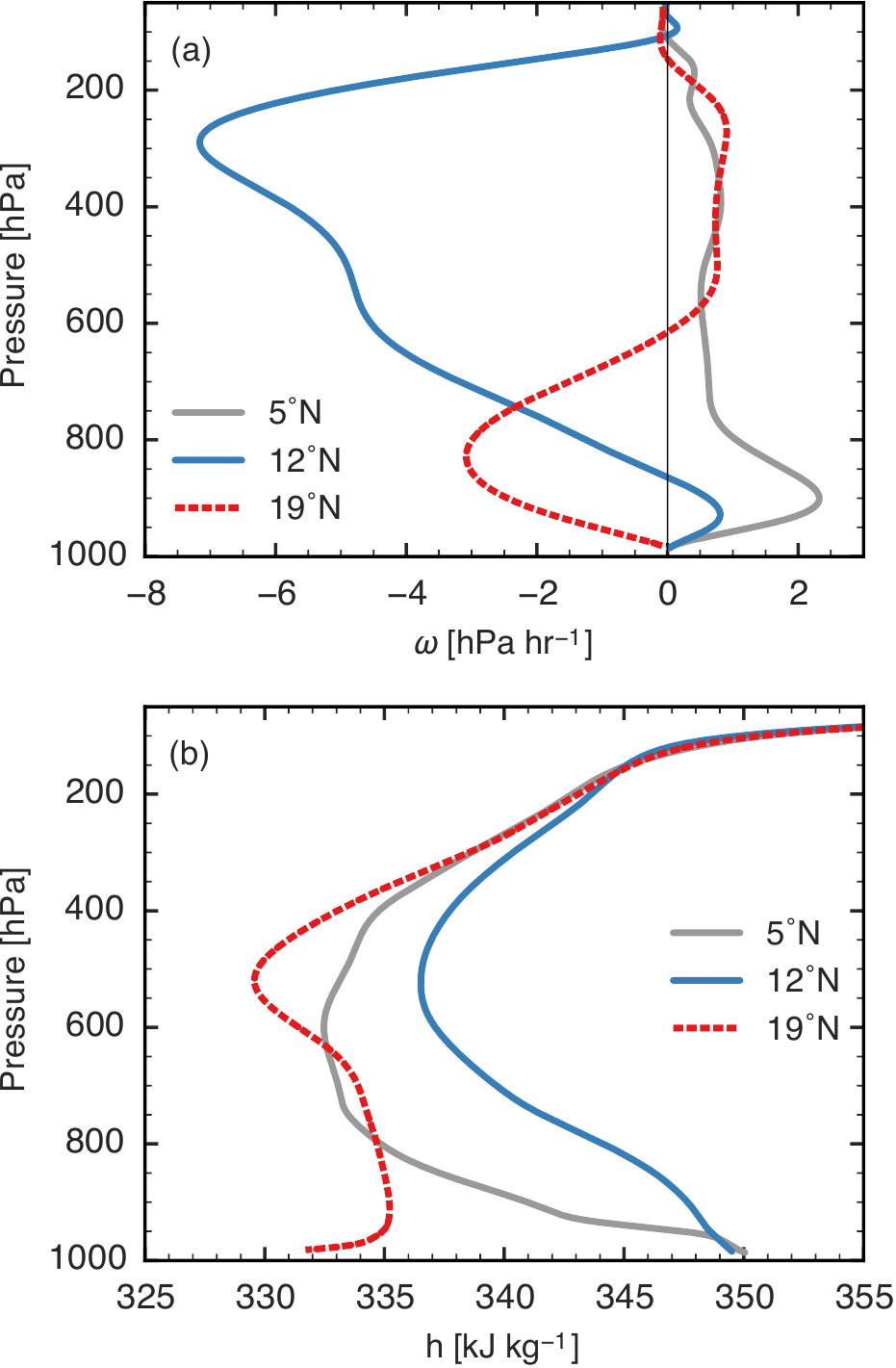}
    \caption{(a) Time mean pressure velocity [hPa hr$^{-1}$] and (b) $h$ at the three latitudes indicated, corresponding to the southward $h_b$ maxima (5\dgN), the northward $h_b$ and precipitation maxima (12\dgN), and the peak heat low ascent (19\dgN).}
    \label{fig:omegahprofiles}
\end{figure}

There are large differences in vertical velocity and $h$ in the free troposphere above these two $h_b$ maxima (Fig.~\ref{fig:omegahprofiles}). At 5\dgN, which is the ocean-grassland boundary, slow subsidence occurs above 800~hPa at a rate that produces the adiabatic warming needed to balance clear-sky radiative cooling (note the similar rate of subsidence in the upper troposphere in the subsiding branch of the summer Hadley cell at 19\dgN).  There is much stronger subsidence below 800~hPa at 5\dgN, which we speculate is dynamically forced by low-level air descending after it undergoes the equatorial jump.  In contrast, at 12\dgNs near the ITCZ, top heavy ascent characteristic of deep convecting regions is seen above 850~hPa, along with weak low-level subsidence that is the time mean of strong ascent during episodes of precipitation and descent at other times in the subsiding branch of the desert's shallow meridional circulation.  Above the subcloud layer, $h$ is substantially lower at 5\dgNs than at 12\dgNs (Figs.~\ref{fig:sfnuh}c, \ref{fig:omegahprofiles}b).  The largest vertically averaged $h$ is at 12\dgN, nearly coincident with the continental ITCZ, and there is a secondary maximum in vertically averaged $h$ on the equator coincident with the near-equatorial precipitation peak (Figs.~\ref{fig:sfnuh}c, d).  Vertical $h$ gradients are small over the desert, as expected due to vertical mixing by dry convection \citep[e.g.][]{Marsham2013,Birch2014}. 

We suggest that the propensity for nearly all moist convective plumes to entrain free tropospheric air \citep[e.g.][]{Romps2010} makes $h$ above the subcloud layer relevant to CQE diagnostics of ITCZ location.  Although $h_b$ is slightly larger at 5\dgNs than at 12\dgN, $h$ in the middle to lower free troposphere is much larger at 12\dgN.  In fact, between 900 hPa and 300 hPa there is a local minimum in $h$ at 5\dgN.  Any entraining convective updraft will thus be more inhibited at 5\dgNs and, even if it did extend to the upper troposphere, would achieve neutral buoyancy with cooler free-tropospheric temperatures.  So instead of treating $h_b$ as our primary thermodynamic variable, we use $h$ averaged over the lower troposphere (from 20 hPa above the surface to 500~hPa), $h_\mathrm{lower}$.  While there are many ways to take a weighted average of $h$ to account for convective entrainment \citep[e.g.][]{Holloway2009}, here we opt for a simple mass-weighted average over this fixed layer.  Indeed, the maxima in $h_\mathrm{lower}$ and upper-tropospheric $h^*$ are collocated and nearly equal in magnitude at 12\dgN, and $h_\mathrm{lower}$ has a local minimum, rather than a global maximum, at 5\dgNs (Fig.~\ref{fig:cqeeb}).  There is a secondary maximum of $h_\mathrm{lower}$ in the center of the near-equatorial precipitation peak, but $h_\mathrm{lower}$ at the equator is about 5 kJ kg\mones less than it is at 12\dgN.  Given the comparatively weak meridional gradient in $h^*$, the equatorial atmosphere is then expected to be more convectively stable than the continental ITCZ, and free-tropospheric temperatures are expected to be set dynamically outside the ITCZ by the cross-equatorial Hadley circulation \citep[e.g.][]{Emanuel1995}.  Using $h_\mathrm{lower}$ as our low-level thermodynamic variable provides a modified CQE estimate $\hat\varphi_\mathrm{CQE:DEEP}$ that is only one degree poleward of the ITCZ. 

Precipitating convection in our model thus seems to be fairly sensitive to free-tropospheric humidity (and consequently $h$), in contrast to many global models with $O$(100~km) resolution \citep[e.g.][]{Derbyshire2004}.  This sensitivity may arise from the moderately high resolution of our model (15 km horizontal grid spacing), permitting some mesoscale convective organization, as well as a convective parametrization that incorporates a representation of enhanced convective entrainment. The Kain-Fritsch scheme \citep{Kain1990,Kain2004} is a mass flux parameterization in which convective updrafts are represented as an entraining-detraining plume.  Lateral entrainment is represented by creation of an ensemble of mixtures of in-cloud and environmental air which are then sorted by buoyancy, with positively buoyant parcels continuing upward in the plume and negatively buoyant parcels detraining.  Older versions of the parametrization produced an increase in cloud top height as environmental humidity decreased, counter to the sensitivity seen in cloud resolving models; that bias was attributed to insufficient entrainment \citep{Jonkers2005}. Modern forms of the Kain-Fritsch scheme, including the version in the WRF model version 3.3, require the mixing that occurs before the buoyancy sorting to incorporate at least 50\% environmental air \citep{Kain2004, DeRooy2013}.  Although biases in how the Kain-Fritsch scheme represents entrainment surely exist \citep[e.g.][]{DeRooy2013}, the relatively high resolution of our model and the enforced minimum entrainment in its convection scheme likely enhance its sensitivity to free-tropospheric humidity. 

\begin{figure}
    \centering
    \includegraphics[width=19pc]{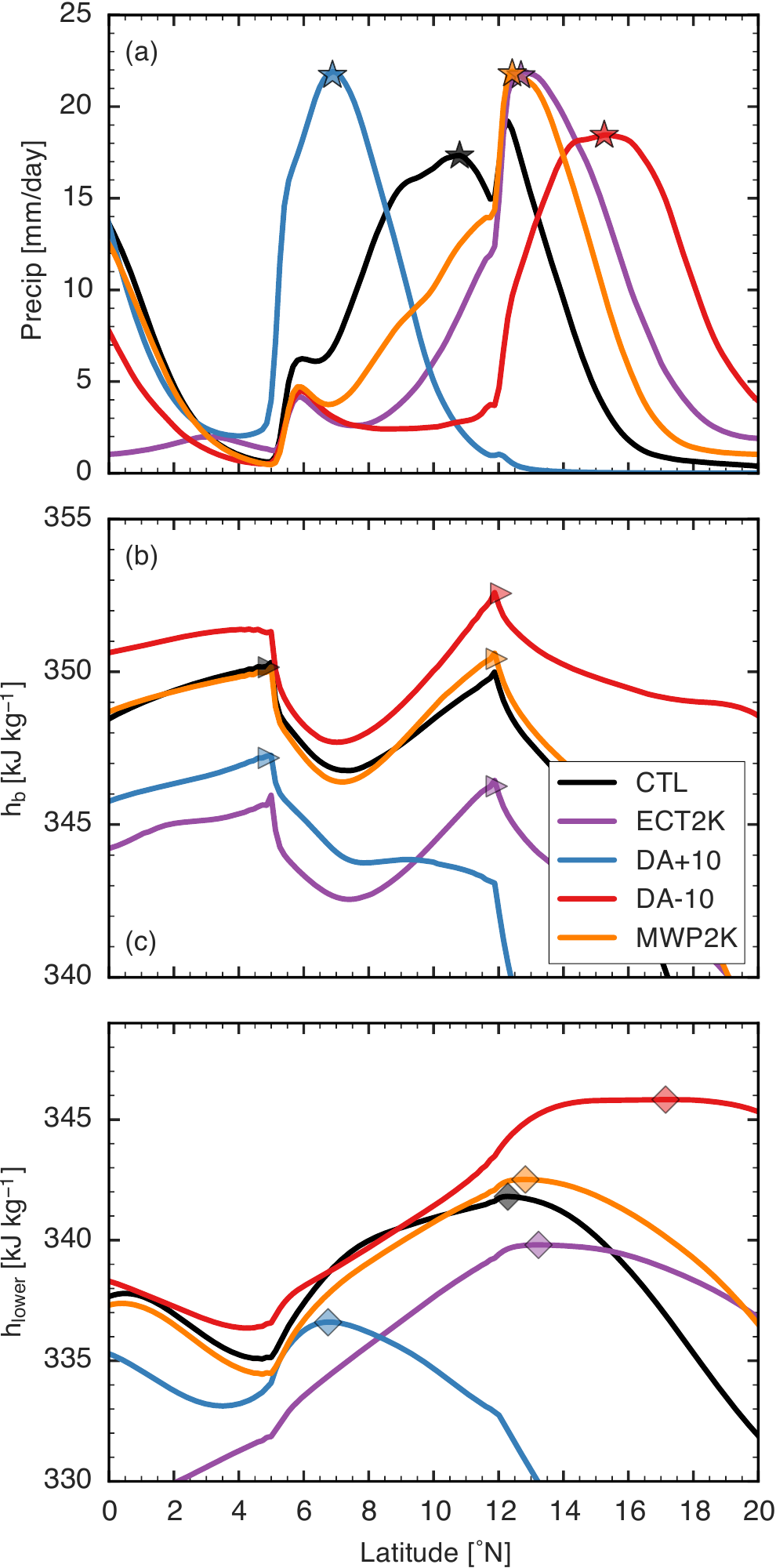}
    \caption{(a) Precipitation with $\phi_\mathrm{ITCZ}$ marked. (b) $h_b$ with $\hat\varphi_\mathrm{CQE:BL}$ marked. (c) $h_\mathrm{lower}$ with $\hat\varphi_\mathrm{CQE:DEEP}$ marked. Colors indicate model integration according to legend. }
    \label{fig:cqecomparison}
\end{figure}

We now turn our attention to how applied forcings modify the control state. As a representative sample, we consider the control and four other integrations: 2 K equatorial cold tongue (ECT2K), 2 K midlatitude warm pool (MWP2K), and two desert albedo $\pm0.10$ forcings (DA$\pm$10). Previous idealized studies of the Sahel \citep{Peyrille2007} have found the precipitation maximum shifts in response to local and remote forcings, and our results support this claim (Fig.~\ref{fig:cqecomparison}a), as $\phi_\mathrm{ITCZ}$ shifts up to four degrees latitude north and south relative to the control integration. The dual maxima of $h_b$ seen in the control integration are a persistent feature in our model (Fig.~\ref{fig:cqecomparison}b). As in the control integration, $\hat\varphi_\mathrm{CQE:BL}$ can be substantially displaced from the ITCZ. Additionally, the location of the $h_b$ maxima does not covary with the ITCZ, but remain anchored at either the coastline at 5\dgNs or the grassland-desert boundary at 12\dgN. In contrast the $h_\mathrm{lower}$ metric (Fig.~\ref{fig:cqecomparison}c) has a clearer global maximum that better tracks $\phi_\mathrm{ITCZ}$. 

\begin{figure*}
    \centering
    \includegraphics[width=\textwidth]{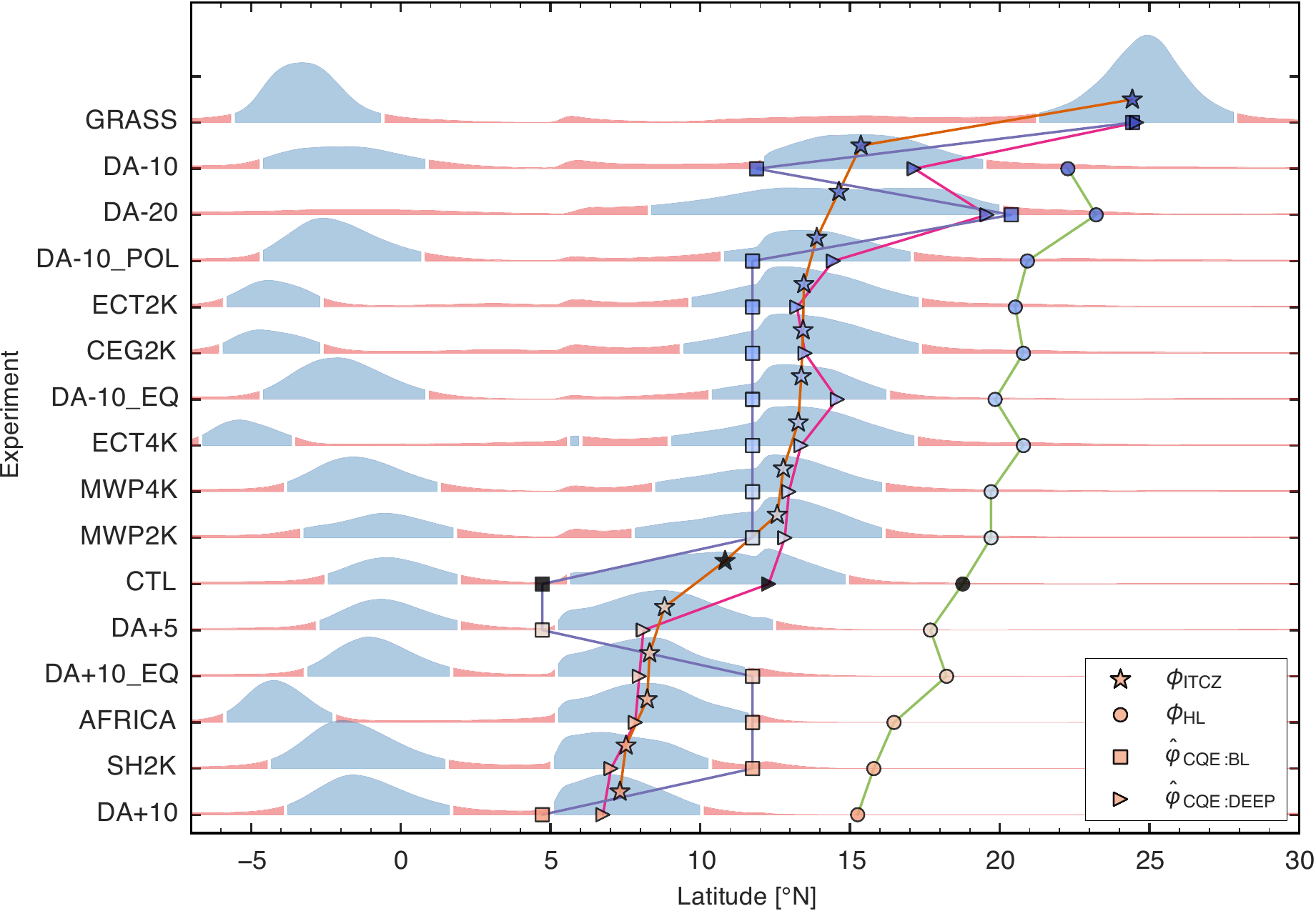}
    \caption{Precipitation distributions for all model integrations shown in shading, with colors identical to Fig.~\protect{\ref{fig:sfnuh}d}. Integrations were sorted according to $\phi_\mathrm{ITCZ}$ (stars). $\hat \varphi_\mathrm{CQE:BL}$ (squares), $\phi_\mathrm{HL}$ (circles), and $\hat \varphi_\mathrm{CQE:DEEP}$ (triangles) are also shown. Marker colors indicate the integration, with bluer colors indicating a more poleward ITCZ, and control integration markers in black.}
    \label{fig:cqepreciptrack}
\end{figure*}

These characteristics of the $h_b$ and $h_\mathrm{lower}$ maxima persist across our entire ensemble of integrations (Fig.~\ref{fig:cqepreciptrack}).  The $\hat \varphi_\mathrm{CQE:BL}$ estimator does not follow $\phi_\mathrm{ITCZ}$ well and remains anchored at an edge of the grassland in most integrations. Even if one more loosely defines the continental convergence zone as the region with precipitation rates larger than 5 mm day\mone, there is a poor correspondence with the $h_b$ maximum.  The $\hat \varphi_\mathrm{CQE:DEEP}$ estimator, which we suggest better represents the effect of entrainment on deep convection and its coupling with upper-tropospheric temperatures, better tracks the ITCZ. 

The improved correspondence between $\hat \varphi_\mathrm{CQE:DEEP}$ and the ITCZ seems to occur because averaging $h$ over a deeper layer better represents the effects of mid-tropospheric dry air south and north of the ITCZ.  Low-$h$ air intrudes into the tropical mid-troposphere from the winter hemisphere, with the minimum $h$ positioned between 600 and 800 hPa (Fig.~\ref{fig:sfnuh}c).  This mid-tropospheric minimum produces the large northward shift in  $\hat \varphi_\mathrm{CQE:DEEP}$ relative to $\hat \varphi_\mathrm{CQE:BL}$ in the CTL and other integrations in the middle rows of Fig.~\ref{fig:cqepreciptrack}.  Low-$h$ air also intrudes into the tropics from the desert in the north, where the minimum $h$ is centered near 500 hPa (Figs.~\ref{fig:sfnuh}c, \ref{fig:omegahprofiles}b); this minimum is responsible for the large southward shift in $\hat \varphi_\mathrm{CQE:DEEP}$ relative to $\hat \varphi_\mathrm{CQE:BL}$ in the SH2K, AFRICA, and DA+10\_EQ integrations shown near the bottom of Fig.~\ref{fig:cqepreciptrack}.  The low-$h$ air intruding from the north consists of dry outflow from the desert's shallow meridional circulation and air that subsided in the summer Hadley cell, the latter having southward and downward components of its velocity (Fig.~\ref{fig:sfnuh}a).  The relatively high altitude of this dry layer motivated our choice of 500~hPa as the upper bound for the averaging in $h_\mathrm{lower}$; choosing a thinner layer that does not include this high altitude dry layer (e.g.\ averaging from 20-150~hPa or 20-350~hPa above the surface) does not provide nearly the same improvement over $\hat \varphi_\mathrm{CQE:BL}$. With these modifications, $\hat \varphi_\mathrm{CQE:DEEP}$ estimates the ITCZ typically within a degree and, more importantly, is better correlated with ITCZ latitude than $\hat \varphi_\mathrm{CQE:BL}$. 

The center of the desert heat low moves poleward as the ITCZ moves poleward in response to forcings that cool the winter hemisphere and warm the northern subtropics (Fig.~\ref{fig:cqepreciptrack}).  Dry, mid-tropospheric outflow from the desert's shallow meridional circulation is stronger when the ITCZ is located closer to the equator, and this shallow circulation and its outflow  vanish in the GRASS integration (where the entire desert is replaced with grassland).  In that integration the maxima of $h_b$ and $h_\mathrm{lower}$ are coincident (top row of Fig.~\ref{fig:cqepreciptrack}) because there is no mid-tropospheric $h$ minimum to create an offset.  We do not present detailed diagnostics of these behaviors of the heat low here, but hope to address them in future work.

\subsection{Energy Budget evaluation}

\begin{figure}
    \centering
    \includegraphics[width=19pc]{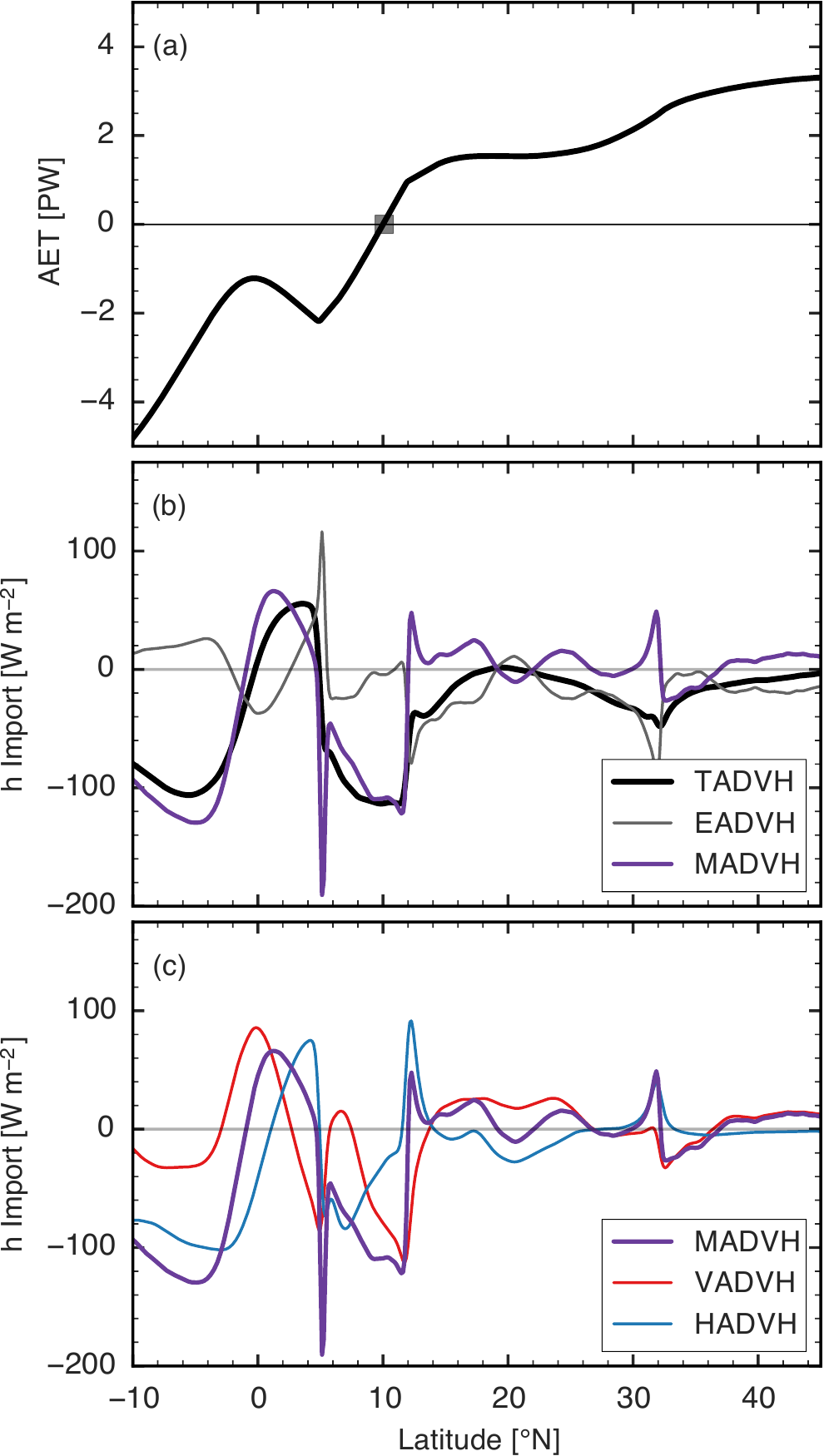}
    \caption{(a) AET for the control integration. (b) Decomposition of the TADVH into mean flow and eddy components MADVH and EADVH respectively. (c) Decomposition of the mean flow MADVH into horizontal and vertical components HADVH and VADVH respectively. Import of energy into the column is considered positive. }
    \label{fig:ctleb}
\end{figure}

Shifting focus to the vertically integrated energy budget, we see that the energy flux equator, $\hat \varphi_\mathrm{EFE}$, lies at 10\dgNs in the control integration and is nearly coincident with the continental ITCZ (Figs.~\ref{fig:ctleb}a, \ref{fig:sfnuh}d).  The circulation diverges energy out of the ITCZ and the entire grassland region (Fig.~\ref{fig:ctleb}b), as expected for the ascending branch of a circulation with a positive GMS.  There are strong contrasts in the vertically integrated energy flux convergence (TADVH) at the edges of the grassland.  In particular, the circulation converges energy into the oceanic region between the equator and 5\dgN, where there is subsidence throughout the troposphere and where surface enthalpy fluxes plus radiation extract energy from the column to balance TADVH.  This region may be analogous to the East Atlantic cold tongue just south of the West African coast, and is notable because it produces a nonmonotonic dependence of AET on latitude that deviates from the linear dependence posited by simple energy budget theories of ITCZ location (e.g. K08, BS14).

Although we do not focus on diagnostics of the GMS, we note that its sign depends on the  definition used.  If the GMS is assumed to be a ratio of AET to the meridional mass flux above 500 hPa, as in \citet{Hill2015}, then the GMS is positive in the domain of the Hadley cell (and is poorly defined, by construction, in the ITCZ).  In contrast, if the GMS is defined as the ratio of TADVH to some weighted average of the upward motion, as in \citet{Peters2008}, then the GMS is positive between the equator and the poleward edge of the continental ITCZ, but is negative in the winter  branch of the Hadley cell. If the GMS is defined as the negative of the ratio of energy flux convergence to moisture flux convergence \citep{Raymond2009}, the region between the equator and the ocean-grassland coast has a negative GMS. And defining the GMS in terms of the vertical advection of $h$ by the time mean $\omega$, as in \citet{Neelin1987} and \citet{Sobel2007}, provides a poor approximation to the actual flow energetics because  mean vertical advection accounts for only a small part of TADVH in many parts of the domain (Fig.~\ref{fig:ctleb}c).  Mean vertical advection is  smaller than mean horizontal advection (i.e. $\left|\mathrm{VADVH}\right| < \left|\mathrm{HADVH}\right|$) in many places, showing that mean horizontal $h$ gradients cannot be neglected in the energy budget.

Decomposition of TADVH into its various components reveals other notable features.  Mean flow advection produces most TADVH in the ITCZ and throughout nearly all of the winter Hadley cell (Fig.~\ref{fig:ctleb}b), but the relative importance of mean horizontal advection and mean vertical advection changes with the region (Fig.~\ref{fig:ctleb}c).  One exception lies at the ocean-grassland boundary, where a pair of oppositely signed peaks in mean and eddy advection (Fig.~\ref{fig:ctleb}b) are associated with a strong land/sea breeze cell  --- mean flow advection is negative there because the mass flux crosses strong horizontal and vertical gradients of $h$ in the lower troposphere (Fig.~\ref{fig:sfnuh}c).  Interestingly, eddy advection produces most TADVH over the desert, from 12-32\dgN, where mean flow advection is of opposite sign to eddy advection and TADVH (Fig.~\ref{fig:ctleb}b).  This likely occurs because subsidence in the summer Hadley cell converges energy into the desert region, primarily in the middle and upper troposphere, while transient eddies balance the net column energy source provided by surface enthalpy fluxes and radiation.  Mass streamlines of the shallow meridional circulation over the desert are largely parallel to $h$ contours, indicating that this circulation contributes little to the vertically integrated energy budget in that region (consistent with the opposing signs of mean horizontal and mean vertical advection there).

\begin{figure}
    \centering
    \includegraphics[width=19pc]{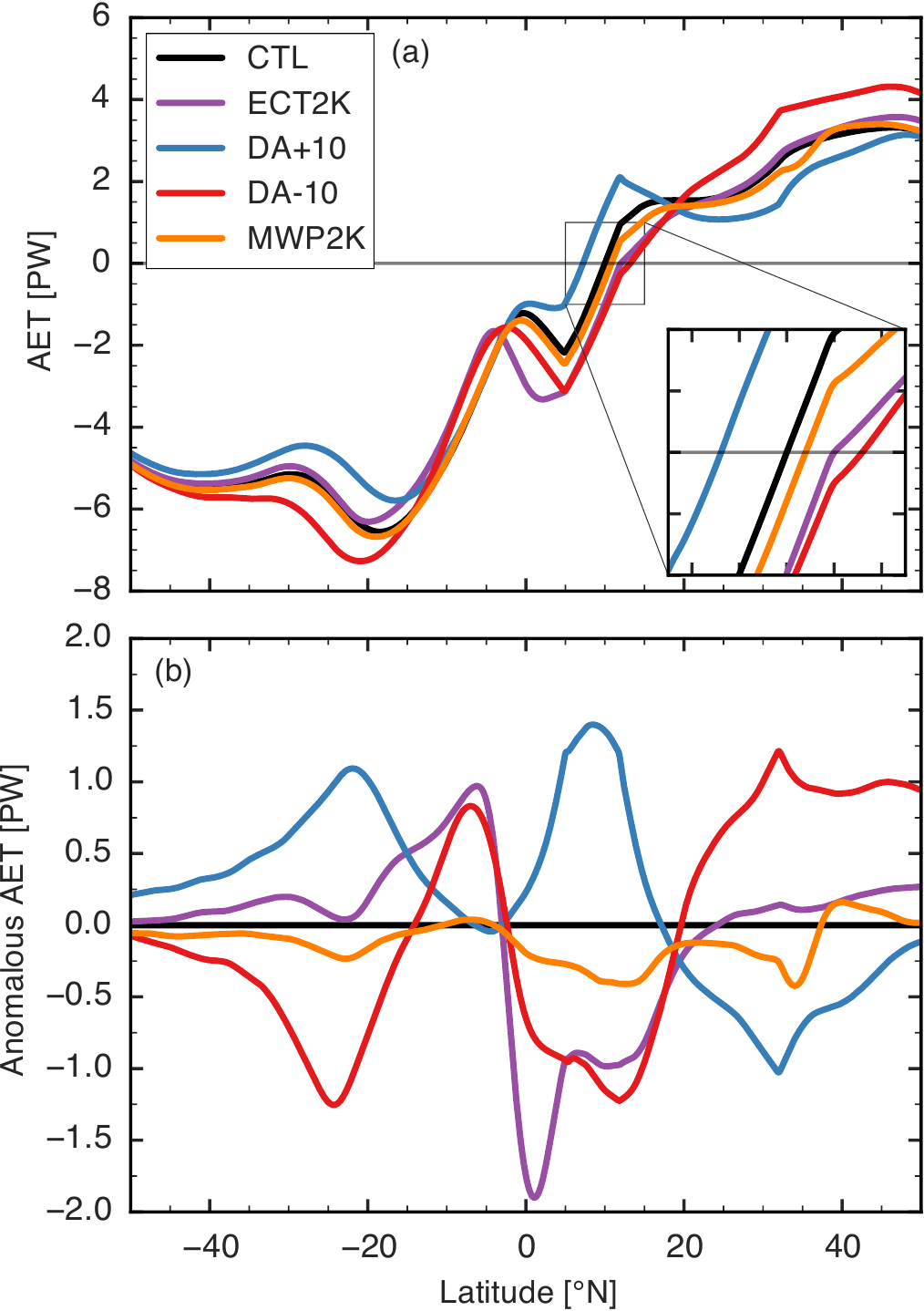}
    \caption{(a) AET for five representative model integrations according to legend. A 3X zoomed inset plot from 5\dgNs to 15\dgNs shows the location of $\hat\varphi_\mathrm{EFE}$. (b) Anomalous AET with respect to control integration. }
    \label{fig:ebaet}
\end{figure}

The northward atmospheric energy transport for the same set of integrations examined in the previous subsection shows that the EFE shifts away from imposed anomalous atmospheric energy sinks and toward imposed sources (Fig.~\ref{fig:ebaet}).  Furthermore, the ITCZ lies within two degrees latitude of the EFE (compare Figs.~\ref{fig:cqecomparison}a and \ref{fig:ebaet}a), so the EFE provides a relatively successful diagnostic of ITCZ position. If we take the control integration EFE latitude to be $E_0$, anomalous energy sinks south of $E_0$ (as in the ECT2K integration) and anomalous sources north of $E_0$ (the MWP2K and DA-10 integrations) produce anomalous southward energy transport across $E_0$ and an associated northward EFE shift.  Similarly, the DA+10 integration exhibits a southward EFE shift compared to the control.  The  anomalous AET near $E_0$ can be used to infer the movement of the EFE because these EFE shifts are generally a few degrees at most. Yet it is clear that the anomalous AET is nonzero far from the region of the imposed forcing.  For example, an increase in desert albedo produces local anomalous energy fluxes into the desert region, but anomalous fluxes on the other side of the equator are nearly as large (blue line in Fig.~\ref{fig:ebaet}b).  

\begin{figure}
    \centering
    \includegraphics[width=19pc]{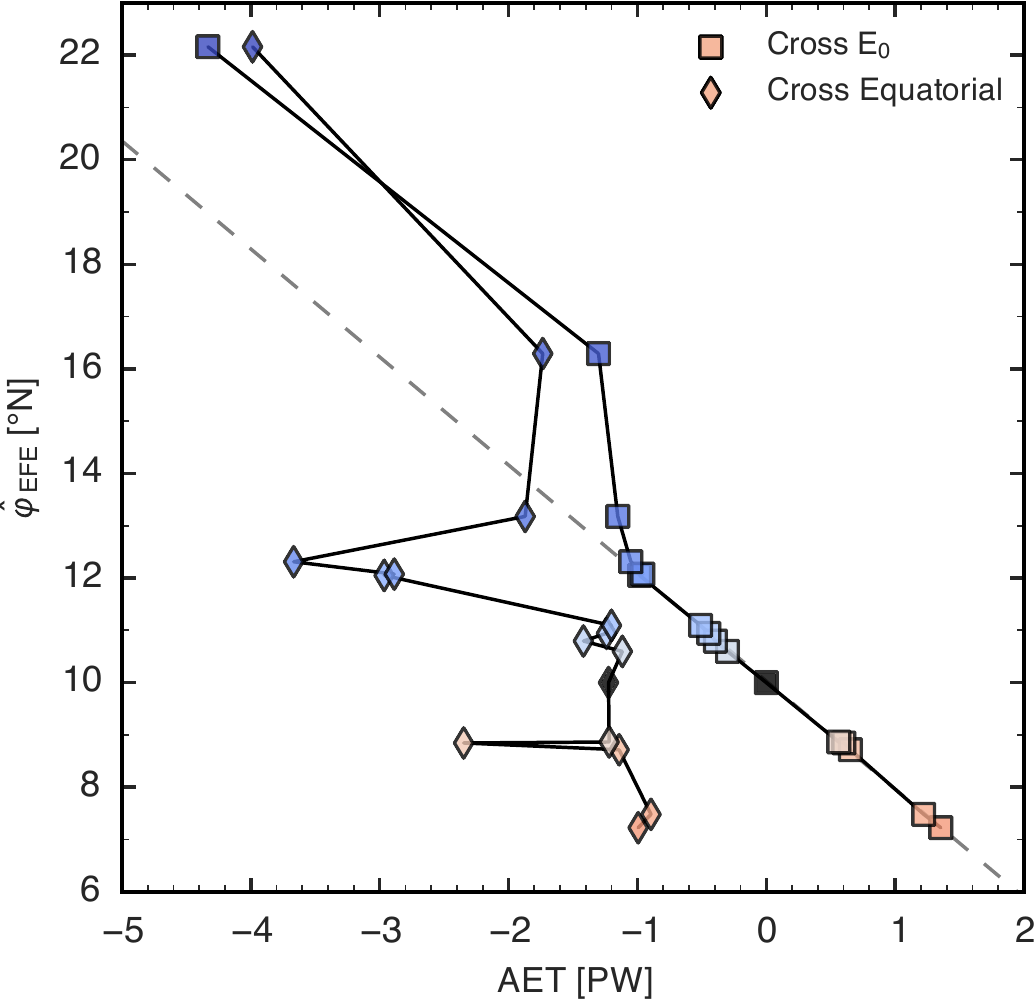}
    \caption{Atmospheric energy transport across the physical equator (diamonds) and across the control integration energy flux equator (squares) at 10\dgNs plotted against $\hat\varphi_\mathrm{EFE}$. Each point represents one model integration, with marker colors as in \protect{Fig.~\ref{fig:cqepreciptrack}}. Linear fit shown (slope = 2.26\dgs PW$^{-1}$) for cross-$E_0$ transport, excluding the three integrations with the largest value of EFE. }
    \label{fig:bs14pred}
\end{figure}

Previous studies have inferred the position of the EFE and ITCZ from the amount of cross-equatorial energy transport (e.g.\ D13), but the non-monotonicity in AET near the equator in our model complicates this sort of estimation. The relationship between EFE latitude and AET at the geographic equator is far from linear (Fig.~\ref{fig:bs14pred}) because our forcings produce large changes in slope of the near-equatorial AET (Fig.~\ref{fig:ebaet}a). However, AET slope near the EFE itself changes little in response to the forcings.  This motivates linearizing the AET about $E_0$ in the control integration (which lies at 10\dgN) to estimate the perturbed EFE location.  This greatly improves a linear estimate of EFE location, which we obtain by fitting all integrations but the three in which the EFE is located furthest poleward.  Deviation from a linear scaling is not surprising for such large shifts, and the remarkably high degree of linearity for smaller shifts indicates that the energy flux divergence in the ITCZ is nearly constant for those smaller shifts.  This linear fit of EFE location to cross-$E_0$ AET yields a slope of 2.26\dgs PW$^{-1}$, roughly consistent with the cross-equatorial value of about 2.4\dgs PW$^{-1}$ obtained by D13 using observations and global models. 

\begin{figure}
    \centering
    \includegraphics[width=19pc]{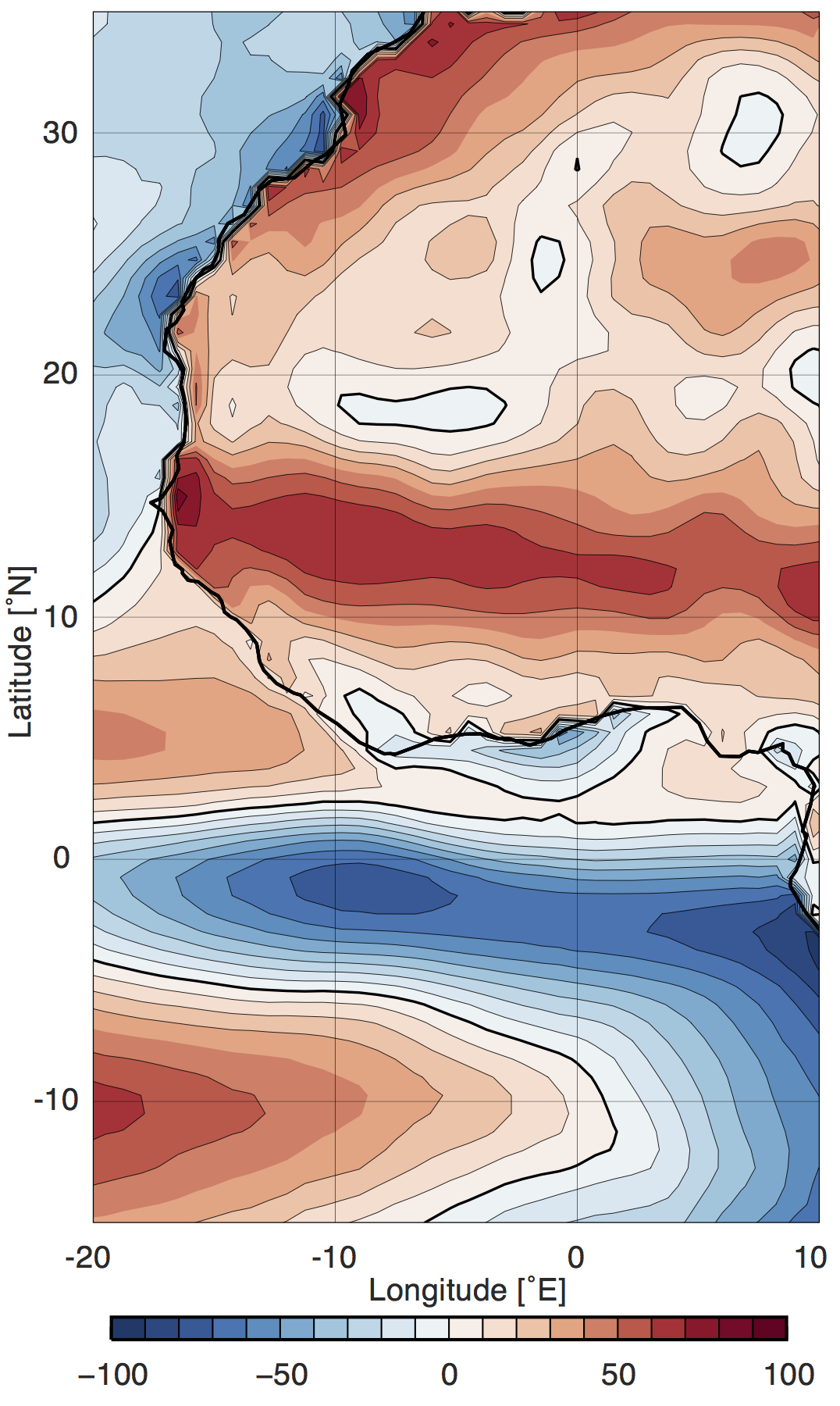}
    \caption{ERA-Interim 1979-2014 JJAS $\overline{\left<Q\right>}$ climatology, in W m\mtwo.}
    \label{fig:erai}
\end{figure}

\begin{figure*}
    \centering
    \includegraphics[width=\textwidth]{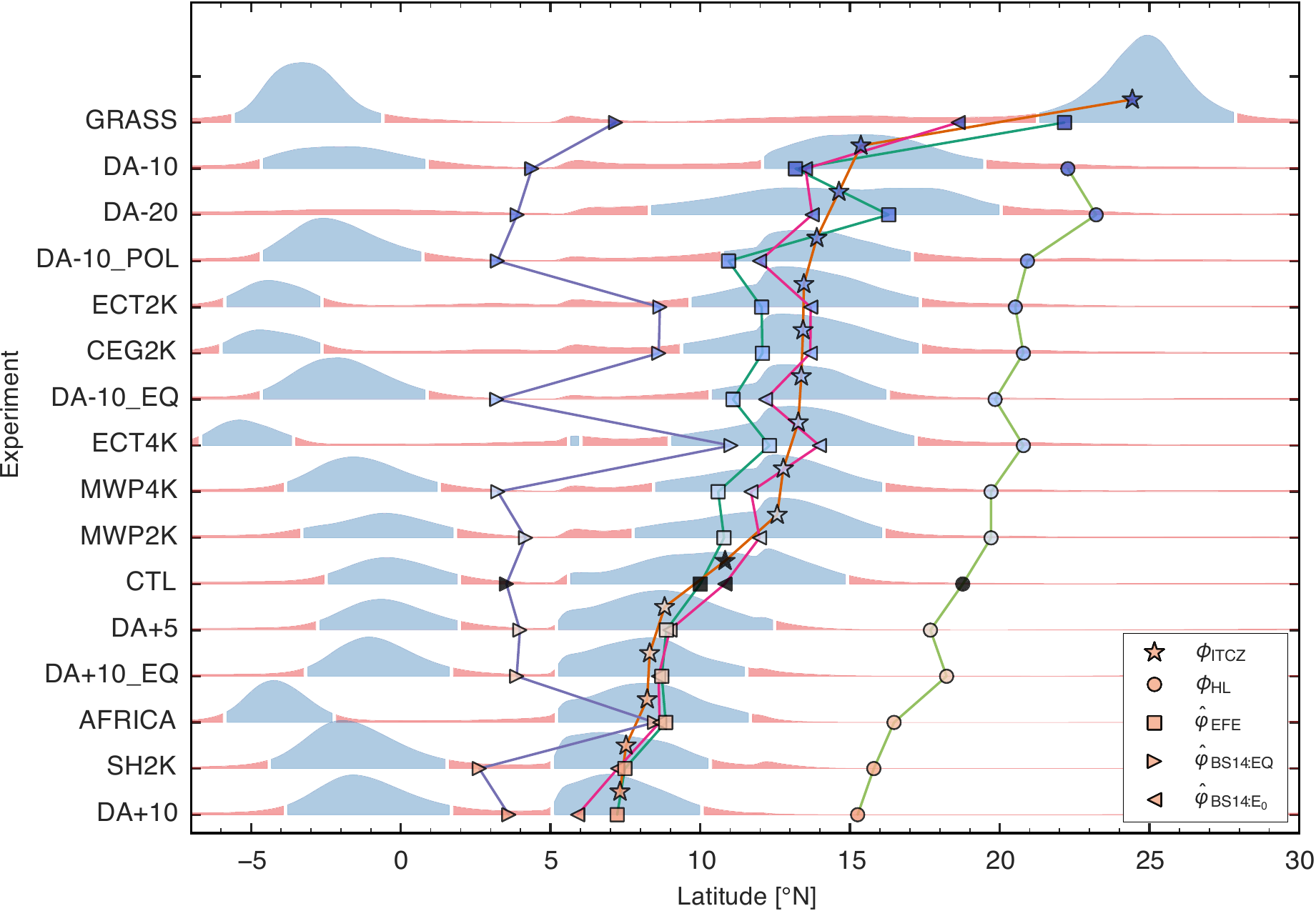}
    \caption{Precipitation distributions as in \protect{Fig.~\ref{fig:cqepreciptrack}}. $\phi_\mathrm{ITCZ}$ (stars), $\hat \varphi_\mathrm{EFE}$ (squares),  $\phi_\mathrm{HL}$ (circles), $\hat \varphi_\mathrm{BS14:EQ}$ (triangle right), and $\hat \varphi_\mathrm{BS14:E_0}$ (triangle left).}
    \label{fig:ebpreciptrack}
\end{figure*}

The AET across $E_0$ is also a successful diagnostic of ITCZ position.  We use the BS14 estimate of ITCZ location linearized about $E_0$, as expressed in (\ref{equ:BS14mod}), but we apply an additional correction $\delta = (\phi_\mathrm{ITCZ})_\mathrm{CTL}-(\hat \varphi_\mathrm{EFE})_\mathrm{CTL}$  to account for the offset between the control integration EFE and ITCZ. We calculate $\delta=0.8^{\circ}$, and denote the corrected estimate of ITCZ latitude as $\hat \varphi_\mathrm{BS14:E_0}$.  This estimate lies within two degrees of the actual ITCZ location for all integrations except the one in which the entire desert was replaced by grassland (GRASS), for which a linear estimate might not be expected to work well (Fig.~\ref{fig:ebpreciptrack}).

The near-equatorial atmospheric energy sink in our model has a real-world analogue that makes the linearization about $E_0$ potentially relevant to observations.  Using the ERA-Interim reanalysis \citep{Dee2011}, we calculated the climatological value of $\overline{\left<Q\right>}$ for June-September 1979-2014 (Fig.~\ref{fig:erai}). The quantities involved are obtained from accumulated 6 to 12-hour forecasts of surface sensible and latent heat fluxes and atmospheric radiative fluxes, so this estimate of $\overline{\left<Q\right>}$ may be more strongly biased than quantities that are directly constrained by assimilated observations.  Nonetheless, this estimate of $\overline{\left<Q\right>}$ shows a large energy sink over the Atlantic equatorial cold tongue that is of a magnitude unlikely to be explained by ERA-Interim bias. A secondary region of negative $\overline{\left<Q\right>}$ over the coastal Gulf of Guinea is likely caused by a local SST minimum created by coastal upwelling.  These features would require the AET to vary non-monotonically with latitude when a limited zonal mean is taken, assuming that  the zonal component of the divergent energy fluxes does not compensate.  Previous studies of the energy budget have not been concerned with this non-monotonicity because they focused on zonal means taken over all longitudes (e.g.\ K08, D13). However, very recent work \citep{Adam2016} explores the energy budget in the limited zonal mean. 

\begin{figure}
    \centering
    \includegraphics[width=19pc]{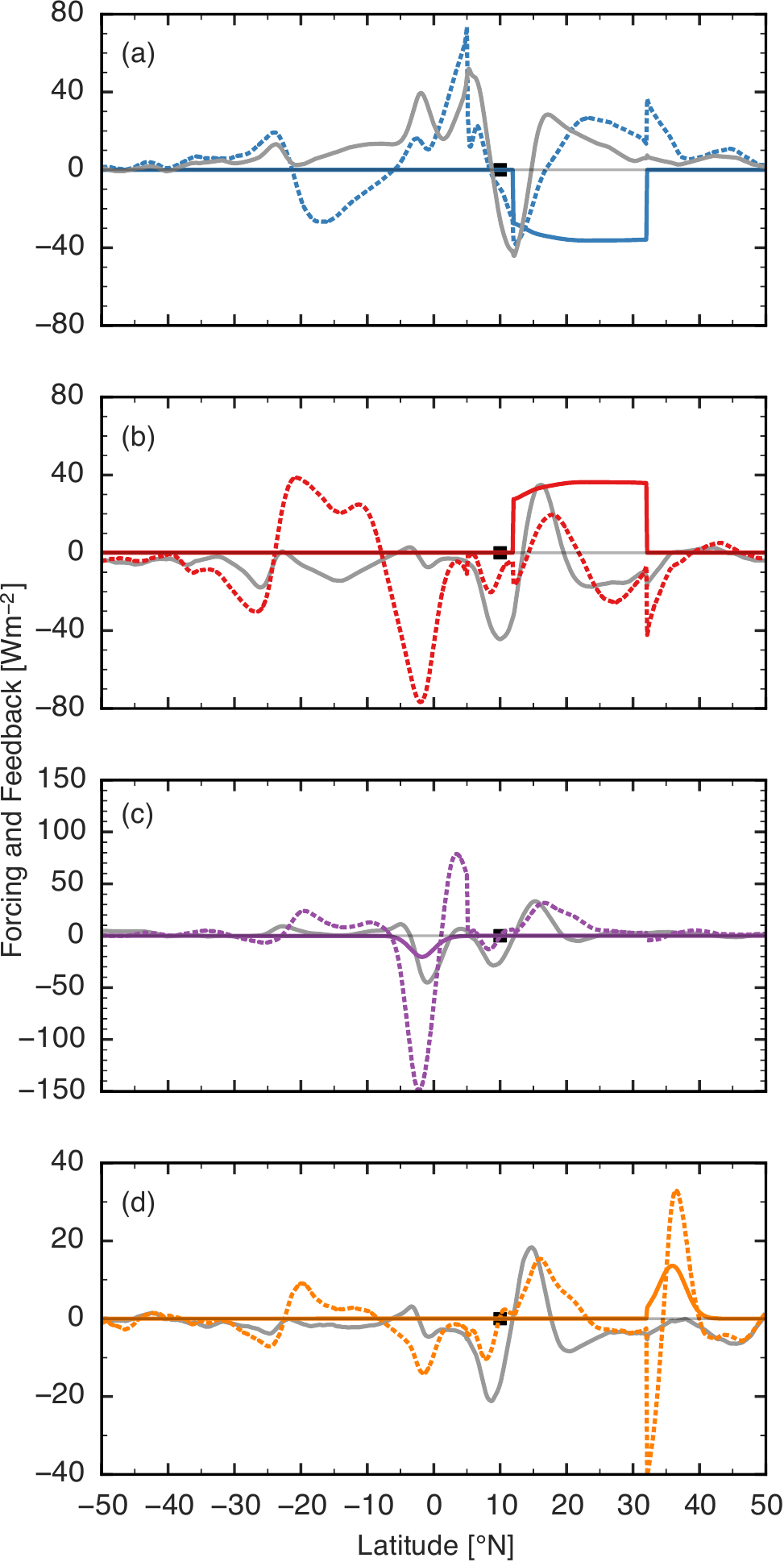}
    \caption{Column integrated longwave heating $\overline{\left<LW\right>}$ (solid gray), forcing (solid colored) and feedback (dotted colored) for four integrations. (a) DA+10 (b) DA-10 (c) ECT2K (d) MWP2K. $E_0$ marked at 10\dgNs for all.}
    \label{fig:forcingfeedback}
\end{figure}

Although the energy budget allows one to correctly guess the sign of the ITCZ shift in response to a forcing, quantitative estimates of ITCZ position based on the energy budget prove to be diagnostic, rather than prognostic, because of large energetic feedbacks.  K08 found that cloud feedbacks greatly altered the ITCZ response to imposed high latitude forcings in an aquaplanet model, and \citet{Seo2014} found that remote high latitude forcings are more effective at perturbing the ITCZ than local tropical forcings due to cloud feedbacks. We also find feedbacks to be highly relevant to our model, and to illustrate this we estimate how our forcings directly and indirectly affect $\overline{\left<Q\right>}$ (recall that energy flux anomalies are related to $\overline{\left<Q\right>}$ anomalies via equations [\ref{equ:budget4}] and [\ref{equ:AET}]).
For example, our albedo forcings directly modify the shortwave radiation absorbed by the surface, which in turn alters the surface fluxes $\overline E$ and $\overline H$ due to the small thermal inertia of land.  Since our albedo forcings are over desert where surface evaporation is small, our albedo forcings primarily modify $\overline H$.  In the absence of feedbacks, and neglecting any modification of the atmospheric shortwave absorption, the direct effect of the albedo forcing $\mathcal{F}$ can be approximated as the control integration downwelling surface shortwave radiation multiplied by the negative of the albedo change, 
\begin{equation}
\mathcal{F} = -(\mathrm{SW}_\mathrm{down,CTL})\Delta \alpha 
\end{equation}
This quantity can be subtracted from the anomalous $\overline{\left<Q\right>}$ to get the feedback, $f$,
\begin{equation}
f = \overline{\left<Q\right>}^{\,\prime} - \mathcal{F}
\end{equation}
For SST forcings, we obtain $\mathcal{F}$ by assuming that the forcing directly modifies only $\overline{E}$; assuming that the surface wind speed and surface air relative humidity do not change, we use a bulk formula to estimate the associated $\overline E$ anomaly. 

Decomposition of $\overline{\left<Q\right>}$ anomalies into a forcing and feedback for four integrations shows that feedbacks are generally large and can be nonlocal (Fig.~\ref{fig:forcingfeedback}). The feedback on the DA+10 forcing has a dipole structure associated with the ITCZ shift (from 12\dgNs to 7\dgN), and is mostly accounted for by the change in column-integrated longwave heating (gray line) caused by the shift of moisture in the ITCZ.  Much of the meridional dipole in longwave heating anomaly near the ITCZ is expected to be associated with changes in cloud radiative forcing, so sensitivity of the cloud forcing to choice of convection scheme might alter the ITCZ response by altering this energetic feedback, as suggested by K08 and \citet{Voigt2014a}.  Over the poleward half of the desert, the feedback negates approximately 60\% of the albedo forcing; a negative feedback is expected because the forcing cools the desert which then emits less longwave, compensating for the effect of the forcing on the net surface radiation.  The cooler atmosphere also emits less longwave to space, which is why the vertically integrated longwave flux divergence opposes the direct albedo forcing.  Similar results are seen for the negative albedo forcing. Perhaps most remarkable are the large $\overline{\left<Q\right>}$ anomalies over the southern ocean with the same sign and nearly the same amplitude as the direct albedo forcing.  The energy budget thus does not respond to a forcing by producing a $\overline{\left<Q\right>}$ anomaly of opposite sign to $\mathcal{F}$. 

For the 2 K equatorial cold tongue (Fig.~\ref{fig:forcingfeedback}c), the magnitude of the feedback is approximately 6 times that of the forcing. This is largely driven by changes in mean surface winds which induce changes in ocean evaporation \citep[e.g.][]{Numaguti1995, Boos2008}, although there is also an increase (50 W m\mtwo) in outgoing longwave radiation caused by a local decrease in column water vapor.  The large subtropical feedback on the 2 K midlatitude warm pool forcing (Fig~\ref{fig:forcingfeedback}d) is also due almost entirely to changes in ocean surface evaporation, but the fact that the mean winds change little (not shown) suggests that these changes are driven by gustiness, perhaps on synoptic scales.  Wind-evaporation feedbacks may be overly strong in models with prescribed SST, but wind-induced changes in upwelling or in dynamical ocean heat transports could complicate matters further \citep[e.g.][]{Webster2003}, and would not be represented in models that used a slab ocean. This makes it difficult for our experiments to either support or provide a counterexample to the idea that high latitude forcings are more effective than tropical forcings at causing shifts in the ITCZ \citep{Seo2014}, as ocean surface energy fluxes can change greatly between our high latitude and low latitude forcings. Although the particular feedbacks in our idealized model may in some cases be artifacts caused by the use of fixed SST, they illustrate how local and non-local interactions with column energy sources can prevent the energy budget from providing a quantitatively prognostic indicator of ITCZ shifts.

Estimates of ITCZ position based on the energy budget perform well across our entire ensemble of model integrations (Fig.~\ref{fig:ebpreciptrack}). The energy flux equator, $\hat\varphi_\mathrm{EFE}$, typically lies within a few degrees of the ITCZ.  Energy transport across the geographic equator is a poor predictor of ITCZ location due to the non-monotonicity in AET, as shown by discrepancies between $\hat\varphi_\mathrm{BS14:EQ}$ and ITCZ latitude.  Using the energy transport across $E_0$ to diagnose ITCZ location (via the $\hat\varphi_\mathrm{BS14:E_0}$ estimator) performs well, even slightly better than the EFE, but part of that gain is due to the applied offset $\delta$=0.8\dgs obtained by diagnosing the difference between  EFE and ITCZ latitudes in the control integration. In all cases the EFE and ITCZ move toward imposed energy sources, but there are unpredictable feedbacks.  For example, the 4 K equatorial cold tongue (ECT4K) integration has an ITCZ located slightly equatorward of the ITCZ in the 2 K equatorial cold tongue (ECT2K) integration, which is surprising given the naive expectation that a -4 K forcing should produce a larger northward ITCZ shift than a -2 K forcing.  All integrations with  negative SST anomalies on the equator (ECT2K, ECT4K, CEG2K, and AFRICA) produce a southward shift of the southern hemisphere precipitation peak and a strong reduction in AET on the geographic equator.  This latter effect inflates the value of $\hat\varphi_\mathrm{BS14:EQ}$ [see equation \eqref{equ:BS14}], providing a fortuitous cancellation of errors that improves the performance of $\hat\varphi_\mathrm{BS14:EQ}$ for those forcings. 

\begin{figure}
    \centering
    \includegraphics[width=19pc]{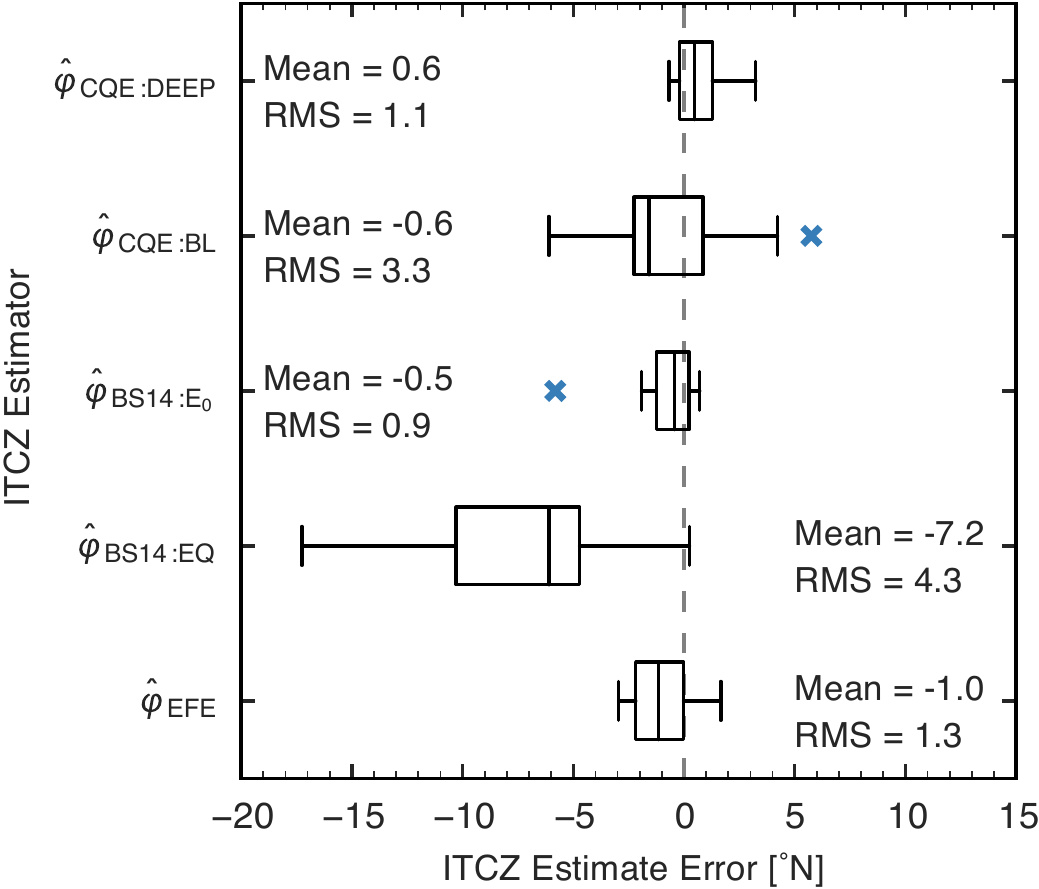}
    \caption{A box and whisker plot of the different ITCZ estimator residuals. The median of the distribution is indicated by the vertical line inside the box. The box width defines the interquartile range (IQR). Outliers (blue crosses) are more than 1.5 IQR from the box boundaries. }
    \label{fig:itczdists}
\end{figure}

When the different estimators of ITCZ location are compared, the one based on energy transports across the geographic equator, $\hat \varphi_\mathrm{BS14:EQ}$, performs worst (Fig.~\ref{fig:itczdists}).  The traditional CQE estimate based on subcloud $h$, $\hat \varphi_\mathrm{CQE:BL}$, is also not effective; it has a large scatter due to the fact that the $h_b$ maxima remain stationary at the edges of the grassland while the ITCZ shifts.  Direct measurement of the EFE performs well, as does the CQE estimate based on $h$ averaged over a deep layer ($\hat \varphi_\mathrm{CQE:DEEP}$) and the AET across 10\dgNs ($\hat \varphi_\mathrm{BS14:E_0}$).  These three estimators are thus good candidates for diagnosing the position of the ITCZ in monsoon regions in the presence of large cross-equatorial energy transports and sources of dry air above the subcloud layer (such as proximal deserts).

\section{Conclusions}
\label{sec:conclusions}

This study examined how well CQE and energy budget frameworks performed in diagnosing the latitude of peak monsoon precipitation in an idealized model.  This beta-plane model represented the shallow meridional circulations associated with the deserts that lie poleward of nearly all observed monsoon regions; it was used to assess the response to an ensemble of idealized  forcings.

We found that the ITCZ shifts toward an imposed atmospheric energy source and away from an energy sink, consistent with previous work \citep[e.g. K08,][]{Chiang2012, Peyrille2007}. The EFE typically lies within two degrees of the ITCZ in our model, providing a good estimate of ITCZ location across a wide range of climates. In contrast, atmospheric energy transport across the geographic equator provided a poor estimate of ITCZ location because of the nonmonotonic dependence of the energy transport on latitude.  This has relevance for the  West African monsoon because an atmospheric energy sink exists over the equatorial East Atlantic (Fig.~\ref{fig:erai}).  If energy transport across a climatological mean ITCZ latitude is instead used, the linear estimate of BS14 describes the ITCZ position accurately. These differences between the geographic equator and the seasonal mean EFE are less pronounced in the global zonal mean, where the ITCZ shifts at most 7\dgs latitude off the equator (D13).  Application of the energy budget framework to regional monsoons is complicated by the fact that the zonal component of  divergent energy fluxes need not be zero when averaging over a limited band of longitudes; nevertheless, our results show that energy fluxes across the geographic equator may be of limited utility for understanding the response of regional monsoons to forcings.

We also tested CQE-based estimates of ITCZ location.  A traditional CQE framework, in which the maximum $h_b$ lies slightly poleward of the ITCZ, failed to capture the ITCZ response to most forcings.  We attributed this failure to the presence of low-$h$ air in the lower and middle troposphere on both sides of the ITCZ:  on the poleward side this dry air originated from the adjacent subtropical desert, and on the equatorward side it originated from the winter hemisphere.  We suggested that this low-$h$ air is entrained into deep convective updrafts so that upper-tropospheric temperatures covary with a weighted average of $h$ over the lower troposphere rather than with $h$ in the subcloud layer.  We modified the CQE framework to account for this entrainment by averaging $h$ over a much thicker layer from 20~hPa above the surface to 500~hPa; the maximum of this $h_{\mathrm{lower}}$ was typically located within a degree of the ITCZ and the maximum upper-tropospheric temperature.

As is common in studies based on idealized models, numerous caveats exist.  Our simulations used perpetual summer insolation, prescribed SST, and prescribed soil moisture.  This is a consistent set of idealizations, since it is problematic to use interactive soil moisture  with perpetual summer insolation \citep{Xie1999}; SSTs would also rise to unrealistic values with perpetual summer insolation unless an oceanic heat sink was prescribed to represent transient heat storage.  Using an annual cycle of insolation with interactive SSTs and soil moisture would have required the model be run at least four times as long, which was not computationally feasible given the fine horizontal resolution (15 km) of our atmospheric model.  We consider this relatively fine resolution to be potentially important in representing the sensitivity of deep convection to entrainment of low-$h$ air above the LCL because it likely allows some mesoscale convective organization, in addition to the minimum entrainment enforced in the Kain-Fritsch convection scheme.  But the main point is that CQE and energy budget frameworks are expected to describe tropical circulations regardless of whether insolation, SST, and soil moisture vary.  The use of fixed SST creates energy sources that would not exist in a model with a slab ocean, but the atmospheric energy budget must still close so that Hadley circulation transports are consistent with ITCZ location.  Furthermore, use of prescribed SST allowed for representation of the atmospheric energy sink associated with the observed equatorial cold tongue in the East Atlantic, which would not exist in a model with an energetically closed slab ocean. 

An equally important caveat is that our zonally symmetric boundary conditions eliminate the possibility that time-mean zonal energy transports influence the position of the ITCZ.  Such transports may be of great importance in setting the response of regional monsoons to imposed forcings, yet these transports vanish in the global zonal mean energy budget frameworks recently used to understand why monsoon precipitation shifts in response to high latitude forcings \citep[e.g.][]{Chiang2012}.  This is less problematic for CQE frameworks, which are based on the local covariance of low-level $h$ and free-tropospheric temperature.

In summary, we demonstrated that deficiencies may exist in the application of a traditional CQE framework and a linear energy budget theory to continental monsoons.  Nonetheless, with  modifications that account for convective entrainment of dry air and the non-monotonicity of meridional atmospheric energy transports, CQE and energy budget theories can accurately describe ITCZ location across a wide range of climate states.  Examining whether these modifications improve diagnostics of ITCZ location in observations and in global models with realistic boundary conditions seems a worthwhile goal of future work.  Unfortunately, neither CQE nor the energy budget are prognostic frameworks due to the feedbacks that alter $h$ and the column-integrated energy source, so a mechanistic and prognostic theory for the response of a regional monsoon to an arbitrary forcing remains elusive. 

\bibliographystyle{ametsoc2014}
 \bibliography{library,custom}
\end{document}